\documentclass[]{aa}  

\usepackage{graphicx}
\usepackage{amsmath,amsfonts,amssymb}
\usepackage{txfonts}
\usepackage{natbib} 
\bibpunct{(}{)}{;}{a}{}{,} % to follow the A&A style

\usepackage{verbatim}
\usepackage{url}
\usepackage{epstopdf}

\begin{document}

   \title{Mixing-length calibration from field stars.} 

   \subtitle{An investigation on statistical errors, systematic biases, and spurious metallicity trends}

   \author{G. Valle \inst{1,2}, M. Dell'Omodarme \inst{1}, P.G. Prada Moroni
     \inst{1,2}, S. Degl'Innocenti \inst{1,2} 
          }
   \titlerunning{Mixing-length - metallicity dependence}
   \authorrunning{Valle, G. et al.}

   \institute{
        Dipartimento di Fisica "Enrico Fermi'',
        Universit\`a di Pisa, Largo Pontecorvo 3, I-56127, Pisa, Italy
        \and
 INFN,
 Sezione di Pisa, Largo Pontecorvo 3, I-56127, Pisa, Italy
 }

   \offprints{G. Valle, valle@df.unipi.it}

   \date{Received 21/12/2018; accepted //}

  \abstract
  % context heading (optional)
 {}
{We critically analysed the theoretical foundation and statistical reliability of the mixing-length calibration by means of standard ($T_{\rm eff}$, [Fe/H]) and global asteroseismic observables ($\Delta \nu$, $\nu_{\rm max}$) of field stars. We also discussed the soundness  of inferring a possible metallicity dependence of the mixing-length parameter from field stars.}
% methods
{We followed a theoretical approach based on mock datasets of artificial stars sampled from a grid of stellar models with a fixed mixing-length parameter 
$\alpha_{\rm ml}$. We then recovered the mixing-length parameter of the mock stars by means of 
SCEPtER maximum-likelihood algorithm.  We finally analysed the differences between the true and recovered mixing-length values quantifying the random errors due to the observational uncertainties and the biases due to possible discrepancies in the chemical composition and input physics between artificial stars and the models adopted in the recovery. 
}
% results heading (mandatory)
{We verified that the $\alpha_{\rm ml}$ estimates are affected by a huge spread, even in the ideal configuration of perfect agreement between the mock data and the recovery grid of models. While the artificial stars were computed at fixed solar-calibrated $\alpha_{\rm ml} = 2.10$, the recovered values had a mean of 2.20 and a standard deviation of 0.52. Then we explored the case in which the solar heavy-element mixture used to compute the models is different from that adopted in the artificial stars. We found an estimated mixing-length mean of $2.24 \pm 0.48$ and, more interestingly, a metallicity relationship in which $\alpha_{\rm ml}$ increases by 0.4 for an increase of 
1 dex in  [Fe/H]. Thus, a simple heavy-element mixture mismatch induced a spurious, but statistically robust, dependence of the estimated mixing-length on metallicity. The origin of this trend was further investigated considering the differences in the initial helium abundance $Y$ -- [Fe/H] -- initial metallicity $Z$ relation assumed in the models and data. We found that a discrepancy between the adopted helium-to-metal enrichment ratio $\Delta Y/\Delta Z$ caused the appearance of spurious trends in the estimated mixing-length values. An underestimation of its value from $\Delta Y/\Delta Z = 2.0$ in the mock data to $\Delta Y/\Delta Z = 1.0$ in the recovery grid resulted in an increasing trend, while the opposite behaviour occurred for an equivalent overestimation. A similar effect was caused by an offset in the [Fe/H] to global metallicity $Z$ conversion. A systematic overestimation of [Fe/H] by $0.1$ dex in the recovery grid of models forced an increasing trend of $\alpha_{\rm ml}$ versus [Fe/H] of about 0.2 per dex. We also explored the impact of some possible discrepancies between the adopted input physics in the recovery grid of models and mock data. We observed an induced trend with metallicity of about $\Delta \alpha_{ml}$ = 0.3 per dex when the effect of the microscopic diffusion is neglected in the recovery grid, while no trends originated from a wrong assumption on the effective temperature scale by $\pm 100$ K. Finally, we proved that the impact of different assumptions on the outer boundary conditions was apparent only in the RGB phase. 
}
% conclusions heading (optional), leave it empty if necessary 
{We showed that the mixing-length estimates of field stars are affected by a huge spread even in an ideal case in which the stellar models used to estimate $\alpha_{\rm ml}$ are exactly the same models as used to build the mock dataset. Moreover, we proved that there are many assumptions adopted in the stellar models used in the calibration that can induce spurious trend of the estimated $\alpha_{\rm ml}$ with [Fe/H]. Therefore, any attempt to calibrate the mixing-length parameter by means of $T_{\rm eff}$, [Fe/H], $\Delta \nu$, and $\nu_{\rm max}$ of field stars seems to be statistically poorly reliable. As such, any claim about the possible dependence of the mixing-length on the metallicity for field stars should be considered cautiously and critically.}


   \maketitle

\section{Introduction}\label{sec:intro}

One of the major and long-standing weaknesses of state-of-the-art
stellar models is the treatment of superadiabatic convection. A precise treatment of external convection would require 3D hydrodynamical calculations which still cannot cover the wide range of physic quantities needed to model stellar populations in our Galaxy. Moreover their results cannot be easily adopted in stellar evolutionary codes, although attempts to implement
 approximations directly based  on 3D simulations in 1D stellar models exist in the literature \citep[e.g.][]{Lydon1992, Ludwig1999, Arnett2015, Arnett2018a}.

The standard treatment, almost universally adopted in 1D stellar
evolution codes, relies on  mixing-length theory (MLT).
In this framework the efficiency of convective transport and stellar structure in the superadiabatic transition layers 
depends on  four free parameters. Three of these are fixed a priori and define the scheme of the MLT \citep[see e.g.][and reference therein]{Salaris2008}. In the following we adopt the  \citet{bohmvitense58} scheme. The last parameter is the so-called mixing-length $l$, 
which is supposed to be proportional to the pressure scale height $H_p$,
i.e. $l= \alpha_{\rm ml} H_p$, where $\alpha_{\rm ml}$ is a non-dimensional free 
parameter to be calibrated. 

An important consequence of such an approach is that neither the
effective temperature nor the radius of stars with a thick outer
convective envelope (i.e. late type stars) can be firmly predicted 
from first principles by current 1D stellar models since they strongly
depend on the calibrated value of $\alpha_{\rm ml}$. 

Usually this parameter is calibrated on the Sun and then the solar calibrated value is adopted for computing stellar models regardless of the mass, evolutionary stage, and chemical composition.
 Nevertheless,
the extrapolation of the solar calibration to different evolutionary phases,
metallicity, and mass ranges has been questioned both on
theoretical and observational grounds. Indeed, 3D simulations of convection
have suggested that the mixing-length changes as a function of stellar luminosity,
gravity, and metallicity \citep{Trampedach2014, Magic2014}. 
Moreover, a growing amount of observations suggest that the adoption of
the solar calibrated $\alpha_{\rm ml}$ does not allow proper modelling of all types of
stars \citep[see e.g][]{Guenther2000, Yildiz2007, Clausen2009, Deheuvels2011, Bonaca2012, Mathur2012, Wu2015, Joyce2018,Joyce2018b}.

Recently, \citet{Tayar2017} claimed that a mixing-length dependent on metallicity,
i.e. $\alpha_{\rm ml} = 0.1612 \times {\rm [Fe/H]} + 1.9037$, is required for stellar models to match
the effective temperatures of over 3\,000 red giant stars in the APOKASC
catalogue \citep{Pinsonneault2014} for which effective temperature, [Fe/H], and
[$\alpha$/Fe] measurements are available. Interestingly, such a dependence on [Fe/H] is much stronger than that predicted by 3D stellar convection simulations 
by \citet{Trampedach2014} and \citet{Magic2014}. Moreover \citet{Tayar2017} did not detect the predicted dependence on the surface gravity.
Furthermore, \citet{Salaris2018} disputed the result by \citet{Tayar2017} pointing
out that any dependence on metallicity of the mixing-length
disappears if the analysis is restricted to stars with $\alpha$-enhancement [$\alpha$/Fe] < 0.08.  

An alternative approach to investigate the possible dependence of
$\alpha_{\rm ml}$ on stellar characteristics is provided by asteroseismology.
A work by \citet{Bonaca2012}, based on a sample of 90 stars with precise asteroseismic and atmospheric measurements in different evolutionary phases suggested the presence of a strong dependence of $\alpha_{\rm ml}$ on [Fe/H]. This analysis was recently extended by \citet{Viani2018}  on a sample of about 450 stars, reaching the same conclusion that a change on $\alpha_{\rm ml}$ of about 0.75 per [Fe/H] dex is needed. This value is much higher than that reported by \citet{Tayar2017}, but the sample of \citet{Viani2018} contains a significant share of main sequence (MS) stars, for which the detected trend is the highest. When restricting the analysis only to the more evolved stars, \citet{Viani2018} have reported a mixing-length trend of about 0.5 per metallicity dex, which is still more than twice the value by \citet{Tayar2017}.

The non-negligible discrepancies between the recent observational calibrated dependence of the mixing-length on the metallicity and the inconsistencies of these values with 3D simulations urge further investigations.

The aim of the present paper is to explore the theoretical foundation of the 
mixing-length calibration by means of standard and global asteroseismic observables 
of field stars. We followed a purely theoretical approach based on mock datasets of artificial stars, sampled from the same grid of stellar models used 
in the calibration procedure. The comparison between the recovered and true 
mixing-length values allowed us to quantify both the random errors due to the observational 
uncertainties and the possible biases. This kind of analysis allows us to explore the 
best possible scenario in which both artificial stars and stellar models are perfectly known a priori. When dealing with real stars, the reliability of the calibration procedure 
can only get worse. 

As shown in detail in the following, even in this ideal case where the artificial stars 
have been computed by keeping  the mixing-length value fixed to the solar value, spurious trends of the recovered $\alpha_{ml}$ as a function of the metallicity can be 
induced by wrong assumptions in the models used in the calibration, either in the 
chemical composition (heavy-element mixture or helium-to-metal enrichment ratio) or input physics (outer boundary conditions, radiative opacity, etc.). 
An example of the relevance of these effects on free parameters calibrations has been shown  by \citet{TZFor} for the estimation of the convective core overshooting parameter from binary stars.  

The work is organised as follows. The estimation method and  grid of adopted stellar models are presented in Sect.~\ref{sec:method}. The results from the various explored scenarios are given in Sect.~\ref{sec:results}. An analysis of the relevance of the chemical inputs assumed in the recovery is presented in Sect.~\ref{sec:feh}. Sect.~\ref{sec:other-trends} addresses the possible effect of a mismatch between the recovery grid and the observations with respect to possible difference in the microscopic diffusion efficiency, boundary conditions, and effective temperature scale. Finally, some conclusions are presented in Sect.~\ref{sec:conclusions}.

\section{Methods}\label{sec:method}

The main aim of this work is to discuss the statistical robustness of the claimed dependence of the mixing-length parameter on metallicity inferred from observations of field stars. In particular, we are interested to assess how well it is possible to recover the mixing-length value starting from standard and global asteroseismic  observables, given the current errors affecting them. The nature of these errors can be statistical, owing to the achievable precision of the instrumentation, but also systematic, due to offset calibrations. 

In the analysis we assume that a mixture of classical and asteroseismic observables are available. In the first class we have the stellar effective temperature and the metallicity [Fe/H], while in the second class we adopt the large frequency separation $\Delta \nu$ and the frequency of maximum oscillation power $\nu_{\rm max}$, that are available for a large set of stars.

We followed a purely theoretical approach in four steps. First, we computed a dense grid of stellar models spanning a wide range of masses, metallicities, helium abundances, and 
$\alpha_{ml}$  values. Second, we built a mock catalogue of artificial stars by sampling them from the grid of models and adding a Gaussian noise to simulate the observational errors. Third, we estimated the mixing-length of the artificial stars by means of SCEPtER \citep{scepter1, eta, bulge, massloss}, a maximum-likelihood pipeline. Finally, we performed a statistical analysis of the estimated mixing-length values with respect to the true value. This procedure allowed us to quantify the minimum random errors and the systematic biases affecting the estimate of  $\alpha_{ml}$ from field stars in this ideal case.

In light of the  warnings raised by \citet{Salaris2018} about a possible influence of the heavy-element mixture, we performed our investigation either assuming an identical mixture both in the mock stars and in the recovery grid of models, but also adopting a different solar heavy-element mixture for the artificial stars.

Sect.~\ref{sec:grids} describes the grid of models and the input physics adopted in its computation. The Monte Carlo design and the statistical methods used for the recovery of stellar parameters given a set of observables are presented in Sect.~\ref{sec:MC}. 

\subsection{Stellar model grid}
\label{sec:grids}

The estimation procedure required a grid of stellar models sufficiently extended to cover the whole parameter space. To this purpose, we computed a large set of stellar models by means of the FRANEC code \citep{scilla2008, Tognelli2011}, in the same
configuration as was adopted to compute the Pisa Stellar
Evolution Data Base\footnote{\url{http://astro.df.unipi.it/stellar-models/}} 
for low-mass stars \citep{database2012}. The evolution is followed from the pre-MS to the onset of the helium flash at the RGB tip.
For the standard grid, we adopted the solar heavy-element mixture by \citet[][hereafter AS09]{AGSS09}, but a second and smaller grid of models was computed using the solar mixture by \citet[][hereafter GS98]{GS98}. 
Atomic diffusion was included, taking into account the
effects of gravitational settling and thermal diffusion with
coefficients given by \citet{thoul94}. 
Outer boundary conditions were determined by integrating the $T(\tau)$ relation by \citet{KrishnaSwamy1966}.

The reference scenario did not account for the radiative acceleration, rotational mixing, and magnetic field, which are affected by somewhat large uncertainties. Furthermore, the effects of mass loss, the uncertainties in the opacities, and equation of state were neglected.  All these quantities and uncertainties have an impact on the stellar evolution \citep[see e.g.][]{incertezze1, Stancliffe2015}, and thus could alter the grid estimates from real-world objects. However, the present study deals with differential effects in an ideal configuration where stellar models and artificial observations perfectly agree with respect to the adopted input physics, therefore their neglect is justified to our purposes. 
Further details on the stellar models can be found in \citet{cefeidi,eta,binary} and references therein.  

The average large frequency spacing $\Delta \nu$ and
the frequency of maximum 
oscillation power $\nu_{\rm max}$ were obtained using the scaling relations from
the solar values \citep{Ulrich1986, Kjeldsen1995} as follows: 
\begin{eqnarray}\label{eq:dni}
\frac{\Delta \nu}{\Delta \nu_{\sun}} & = &
\sqrt{\frac{M/M_{\sun}}{(R/R_{\sun})^3}} \quad ,\\  \frac{\nu_{\rm
                max}}{\nu_{\rm max, \sun}} & = & \frac{{M/M_{\sun}}}{ (R/R_{\sun})^2
        \sqrt{ T_{\rm eff}/T_{\rm eff, \sun}} }. \label{eq:nimax}
\end{eqnarray}
The validity of these scaling relations in the RGB phase has been questioned in recent years \citep{Epstein2014, Gaulme2016, Viani2017}. Although their reliability poses a severe problem whenever adopted for a comparison with real observational data, it is of minor relevance for our aim because we use exactly the same scaling relations to compute $\Delta \nu$ and $\nu_{\rm max}$ in both the artificial stars and the models. 

The grid of models spans the range [0.8, 1.0] $M_{\sun}$, with a step of 0.01 $M_{\sun}$, and covers the 
initial metallicity interval $-0.4 \; {\rm dex} \leq$ [Fe/H] $\leq 0.4$ dex, with
a step of 0.05 dex. The mass range was chosen in a way to avoid models that develop a convective core in the MS evolution. This allowed us to neglect the further degree of freedom due to the poorly constrained  convective core overshooting efficiency. 
For the AS09 solar mixture we computed -- for each metallicity -- models for three different values of the initial helium abundance by following the 
linear relation, 
\begin{equation}
Y = Y_p+\frac{\Delta Y}{\Delta Z} Z,\label{eq:dydz}
\end{equation}
where the primordial abundance $Y_p = 0.2485$ from WMAP
\citep{peimbert07a,peimbert07b} and a helium-to-metal enrichment ratio $\Delta Y/\Delta Z$ = 1, 2, 3, to explore a sensible uncertainty range \citep{gennaro10}, where $\Delta Y/\Delta Z = 2.0$ corresponds to the reference value for the synthetic systems. For the GS98 solar mixture only, models with $\Delta Y/\Delta Z = 2.0$ were computed.
The initial metallicity $Z$ is in turn linked to the initial [Fe/H] and $\Delta Y/\Delta Z$ by the relation
\begin{equation}
Z = \frac{(1-Y_p) \left(\frac{Z}{X}\right)_{\sun}}{10^{-{\rm [Fe/H]}}-\left( 1+\frac{\Delta Y}{\Delta Z}\right) \left(\frac{Z}{X} \right)_{\sun}}. \label{eq:Z}
\end{equation}
The $(Z/X)_{\sun}$ depends on the adopted solar heavy-element mixture. The effect of changing the heavy-element mixture is explored in Sect.~\ref{sec:feh}.

Ultimately, the grid spans a set of 51 different initial chemical compositions. For each mass, metallicity, and initial helium abundance, we computed models for 21 values of the mixing-length parameter $\alpha_{\rm ml}$ in the range
[1.0, 3.0] with a step of 0.1. With the assumed input physics, the solar-calibrated value  was $\alpha_{\rm ml} = 2.1$. The grid resolution is sufficiently high to impact in a negligible way on the estimates. 

\subsection{Mock catalogues and recovery technique}
\label{sec:MC}

Two mock catalogues each of $N = 5\,000$ artificial stars were generated starting from the grids computed with AS09 and GS98 solar mixtures. 
The objects were sampled from tracks at solar-calibrated mixing-length value ($\alpha_{\rm ml} = 2.1$ in both cases) and with helium-to-metal enrichment ratio $\Delta Y/\Delta Z = 2.0$.
A cut-off in the age of the synthetic objects was imposed at 14 Gyr. The mass and evolutionary stage of the objects were randomly selected, respecting the above-mentioned constraints.

As observational constraints in the recovery procedure we adopted the stellar effective temperature, the surface current metallicity [Fe/H], which can be different from the initial one owing to element diffusion, large frequency spacing $\Delta \nu$, and frequency of maximum 
oscillation power $\nu_{\rm max}$. The two datasets were then subjected to artificial Gaussian perturbations, to simulate observational uncertainties. The values of the typical uncertainties were chosen to match the median values of those reported in \citet{Viani2018}, for ease of comparison. Therefore, we adopted 50 K on $T_{\rm eff}$, 0.08 dex in [Fe/H], 2\% in $\Delta \nu$, and 4\% in $\nu_{\rm max}$.
Additional datasets were generated  to investigate the relevance of the individual observational constraints to the final recovered mixing-length values. 

All the perturbed datasets were then subjected to the recovery adopting the well-tested SCEPtER pipeline \citep{scepter1, eta, bulge, massloss}.
We briefly summarise the technique here.

We let $\cal
S$ be a star for which the observational quantities
$q^{\cal S} \equiv \{T_{\rm eff, \cal S}, {\rm [Fe/H]}_{\cal S},
\Delta \nu_{\cal S}, \nu_{\rm max, \cal S}\}$ are available. Then we let $\sigma = \{\sigma(T_{\rm
        eff, \cal S}), \sigma({\rm [Fe/H]}_{\cal S}), \sigma(\Delta \nu_{\cal S}),
\sigma(\nu_{\rm max, \cal S})\}$ be the observational uncertainty. 

For each point $j$ on the estimation grid of stellar models, 
we define $q^{j} \equiv \{T_{{\rm eff}, j}, {\rm [Fe/H]}_{j}, \Delta \nu_{j},
\nu_{{\rm max}, j}\}$. 
We let $ {\cal L}_j $ be the likelihood function defined as
\begin{equation}
{\cal L}_j = \left( \prod_{i=1}^4 \frac{1}{\sqrt{2 \pi} \sigma_i} \right)
\times \exp \left( -\frac{\chi^2}{2} \right)
\label{eq:lik}
,\end{equation}
where
\begin{equation}
\chi^2 = \sum_{i=1}^4 \left( \frac{q_i^{\cal S} - q_i^j}{\sigma_i} \right)^2.
\end{equation}

The likelihood function is evaluated for each grid point within $3 \sigma$ of
all the variables from $\cal S$. We let ${\cal L}_{\rm max}$ be the maximum value
obtained in this step. The estimated stellar quantities are obtained
by averaging the corresponding quantity of all the models with likelihood
greater than $0.95 \times {\cal L}_{\rm max}$.
Informative priors can be inserted as a multiplicative factor
in Eq.~(\ref{eq:lik}), as a weight attached to the
grid points.

\begin{table*}[ht]
        \centering
        \caption{Datasets adopted in the paper.} 
        \label{tab:datasets}
        \begin{tabular}{p{3.2cm}p{12.8cm}}
                \hline\hline
                label & notes\\ 
                \hline
                AS09 &  Mock data sampled from the \citet{AGSS09} solar mixture grid; recovery with the \citet{AGSS09} solar mixture.\\ 
                GS98 & Mock data sampled from the \citet{GS98} solar mixture grid; recovery with the \citet{AGSS09} solar mixture.\\ 
                AS09-Mp, AS09-Mm &  Same as AS09, but with an offset of $\pm 0.1$ dex in the mock data [Fe/H].\\ 
                AS09-Y1, AS09-Y3 &  Same as AS09, but with $\Delta Y/\Delta Z = 1.0$ and 3.0 in the recovery grids.\\
                GS98-Y1, GS98-Y3 &  Same as GS98, but with $\Delta Y/\Delta Z = 1.0$ and 3.0 in the recovery grids.\\
                AS09-nd & Same as AS09, but reconstructing data on a grid which neglects the effect of microscopic diffusion in the surface [Fe/H].\\
                AS09-bc & Mock data sampled from a grid with boundary conditions from atmosphere tables; recovery as in the AS09 scenario. \\
                AS09-Tp, AS09-Tm & Same as AS09, but with an offset of $\pm 100$ K in the effective temperature of mock data.\\
                \hline
        \end{tabular}
\end{table*}

The recovery was conducted for several different scenarios. In particular the recovery was performed either assuming a perfect match between data and grid (scenario AS09: $\Delta Y/\Delta Z = 2.0$ and AS09 mixture in both cases; Sect.~\ref{sec:as09}), and assuming possible mismatches in the solar heavy-element mixture (scenario GS98; Sect.~\ref{sec:gs98}). Systematic  differences in the helium-to.-metal enrichment ratios between data and grids of models were also explored  (scenarios AS09-Y1, AS09-Y3, GS98-Y1, and GS98-Y3; Sect.~\ref{sec:bias-dydz}). We investigated  the relevance of adopting a wrong scale for the surface [Fe/H] by displacing the data with respect to the grid (scenarios AS09-Mp, AS09-Mm; Sect:~\ref{sec:feh-scale}). Differences between the artificial data and the grid with respect to the efficiency of the microscopic diffusion, boundary conditions, and effective temperature scale were also considered (scenarios AS09-nd, AS09-bc, AS09-Tp, AS09-Tm; Sect.~\ref{sec:other-trends}). The adopted scenarios are listed in Tab.~\ref{tab:datasets}. Each scenario is fully described the first time it is mentioned during the analysis. 
 
\section{Mixing-length parameter estimates}
\label{sec:results}

The values of the mixing-length parameter estimated in the AS09 and GS98 scenarios introduced in Sect.~\ref{sec:MC} served us as starting point for all other investigations and are discussed in detail in the following subsections.

\subsection{AS09 scenario: Ideal agreement between models and data}\label{sec:as09}

The AS09 scenario describes the ideal configuration in which the grid of models are in perfect agreement with the data aside from the unavoidable presence of random observational uncertainties. In fact, in this scenario the mock stars were sampled from the same grid of models used in the recovery procedure. In particular they share the same heavy-element mixture, i.e. \citet{AGSS09} , and the same $\Delta Y/\Delta Z = 2.0$. 

Such a scenario thus allows us to establish the minimum biases and possible trends. When dealing with real stars, a much worse performance must be expected as a consequence of the still present systematic uncertainties affecting stellar models.

\begin{figure*}
        \centering
        \includegraphics[height=16.0cm,angle=-90]{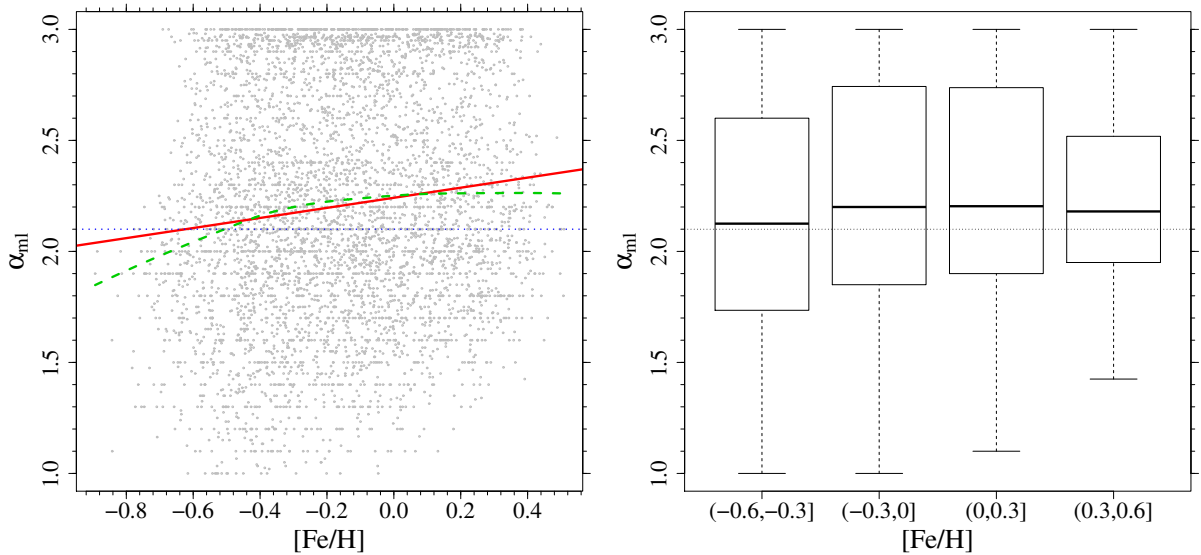}
        \caption{{\it Left}: Scatterplot of the estimated mixing-length values in the scenario AS09. The red solid line shows a linear regression model, while the dashed green line indicates a LOESS smoother of the data (see text). The dotted horizontal line indicates the true value of $\alpha_{\rm ml}$. {\it Right}: Boxplot of the recovered mixing-length values in the scenario AS09 in the adopted metallicity bins.}
        \label{fig:alpha-feh-as09}
\end{figure*}

Fig.~\ref{fig:alpha-feh-as09} shows the estimated values of $\alpha_{\rm ml}$ versus the observed metallicity [Fe/H]. A huge spread of the estimates is apparent: while the true value was $\alpha_{\rm ml} = 2.1$ the recovered values practically cover the entire explored range from 1 to 3. Moreover, the results show a bias towards overestimation; the $\alpha_{\rm ml}$ recovered mean value is 2.20 with standard deviation 0.52. 

It is worth noting that the large spread of the estimated values for the AS09 scenario is only caused by the presence of random observational errors, implying that the adopted observable constraints and the assumed observational errors do not provide a firm ground for the calibration of the mixing-length parameter. A similar conclusion has been presented by \citet{overshooting, BinTeo} in the framework of calibration of the convective core overshooting parameter.

A reduction of the uncertainties in the observational constraints ($T_{\rm eff}$, [Fe/H], $\Delta \nu$, and $\nu_{\rm max}$) can improve the situation, but -- as shown in Sect.~\ref{sec:errori} -- even an overoptimistic shrink of the errors to one-quarter of the previously adopted values  reduces the standard deviation on $\alpha_{\rm ml}$ by only a factor of two.

Aside from the huge variance of the estimated values of $\alpha_{\rm ml}$, a clear trend is evidenced when attempting a linear regression of the mixing-length parameter versus [Fe/H] (straight line in Fig.~\ref{fig:alpha-feh-as09}). The regression coefficient was $0.23 \pm 0.03$.
However, a linear regression is actually not theoretically grounded in this case for several reasons. First, the iron abundances adopted as regressors are affected by non-negligible uncertainties. Second, the values of [Fe/H] are far from being evenly spread in the whole range. Moreover, stellar evolutionary phases are differently represented in different metallicity ranges. This is particularly evident for stellar models with [Fe/H] below $-0.4$, which is the lowest initial [Fe/H] value available in the recovery grid.  This metallicity range is populated by stars whose surface [Fe/H] is depleted by the effect of the microscopic diffusion during the MS evolution. It is therefore clear that this bin lacks of star in the first evolutionary stages and contains a great share of stars that are near the 85\% of their MS evolution, where the effect of the microscopic diffusion on the surface metallicity is the highest. 

\begin{figure}
        \centering
        \includegraphics[height=8.2cm,angle=-90]{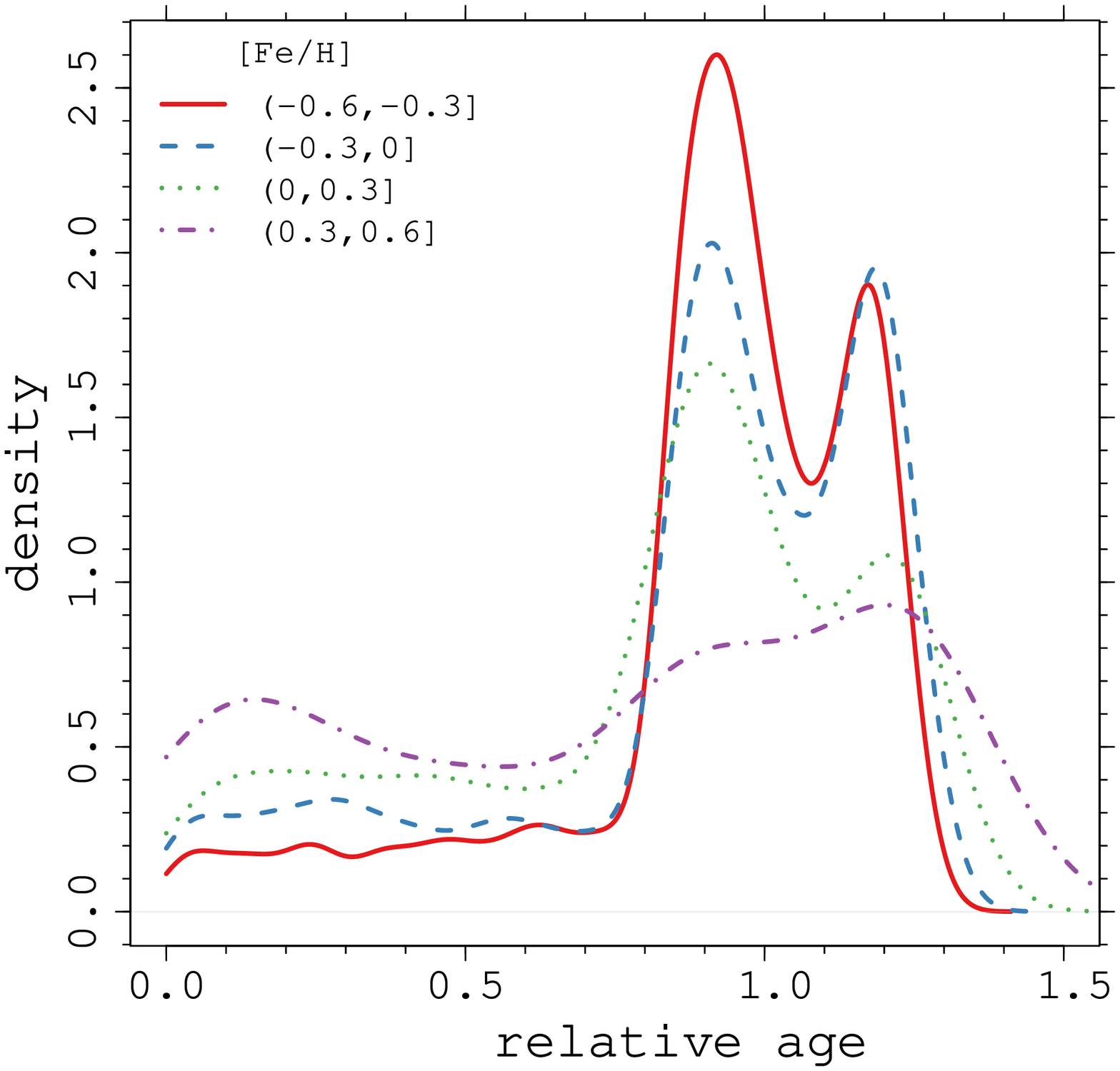}
        \caption{Kernel density of the relative age of the stars in the considered metallicity bins.}
        \label{fig:density-pcage}
\end{figure}

The violation of both the mentioned theoretical assumptions suggests that a linear regression cannot be appropriate and different methods can perform better to explore the existing trends. To this purpose a first degree LOESS smoother was adapted to data\footnote{A LOESS (LOcal regrESSion) smoother is 
a non-parametric  locally weighted polynomial regression technique that is
often used to highlight the underlying trend of scattered data \citep[see       e.g.][]{Feigelson2012,venables2002modern}.}, as is shown by a curved line in Fig.~\ref{fig:alpha-feh-as09}. The LOESS smoother shows substantial differences with respect to the linear fit. A steep increase in the mixing-length value with the metallicity is evident for [Fe/H] $\leq -0.4$ (caused by differences in the represented evolutionary stages, see later), followed by a plateau. However even this method has its drawback because it adopts a moving window to smoothen the trend; in this particular case a fraction $f=0.6$ of data are used for every smoother point. Therefore the results at a given metallicity are influenced by those at different [Fe/H].

\begin{table}[ht]
        \caption{Median of the estimated mixing-length values in the considered metallicity bins, for all the considered scenarios. }
        \label{tab:med-alpha} 
        \centering
        \begin{tabular}{lrrrr}
                \hline
                Scenario & (-0.6,-0.3] & (-0.3,0] & (0,0.3] & (0.3,0.6] \\ 
                \hline
                AS09 & 2.12 & 2.20 & 2.20 & 2.18 \\ 
                GS98 & 2.10 & 2.20 & 2.34 & 2.47 \\ 
                AS09-Y1 & 2.20 & 2.39 & 2.60 & 2.63 \\ 
                AS09-Y3 & 2.05 & 2.05 & 1.93 & 1.90 \\ 
                GS98-Y1 & 2.20 & 2.38 & 2.69 & 2.91 \\ 
                GS98-Y3 & 2.00 & 2.07 & 2.08 & 2.10 \\ 
                AS09-Mp-Y1 & 2.49 & 2.67 & 2.82 & 2.70 \\ 
                AS09-Mp-Y2 & 2.36 & 2.49 & 2.44 & 2.20 \\ 
                AS09-Mp-Y3 & 2.25 & 2.27 & 2.12 & 1.85 \\ 
                AS09-Mm-Y1 & 2.01 & 2.10 & 2.35 & 2.53 \\ 
                AS09-Mm-Y2 & 1.95 & 1.96 & 2.03 & 2.14 \\ 
                AS09-Mm-Y3 & 1.88 & 1.82 & 1.80 & 1.85 \\ 
                AS09-Tp & 2.36 & 2.42 & 2.43 & 2.39 \\ 
                AS09-Tm & 1.92 & 2.00 & 2.02 & 2.00 \\ 
                AS09-nd & 1.90 & 1.98 & 2.05 & 2.20 \\ 
                \hline
        \end{tabular}
\end{table}

A more robust -- while very conservative -- method is to present a boxplot\footnote{A boxplot is a convenient way to summarise the 
variability of the data; the black thick lines show the median of the dataset, while the box indicates the interquartile range; i.e., it extends form the 25th to the 
75th percentile of data. The  whiskers extend from the box until the
extreme data.}  of the mixing-length values binned in metallicity (right panel in Fig.~\ref{fig:alpha-feh-as09}). The plot gives evidence of a nearly stationary behaviour, with the above noticed  drop in the lowest metallicity bin. The median values of $\alpha_{\rm ml}$ in the metallicity bins are reported in Tab~\ref{tab:med-alpha}.

This drop of $\alpha_{\rm ml}$ at low [Fe/H] is related to the differences in the evolutionary stages represented in the four considered metallicity bins, as shown in Fig.~\ref{fig:density-pcage}. The highest metallicity bin shows a notable proportion of models in the first 50\% of the MS evolution and a lack of near turn-off models. This is clearly due to an edge effect, an ubiquitous behaviour in grid recovery algorithms. More in detail, during the MS evolution, the surface [Fe/H] decreases due to diffusion thus depleting the bin which -- at variance with metallicity poor bins -- cannot be populated by the models starting at higher metallicity. Other effects specifically linked to the evolutionary phase are discussed in Sect.~\ref{sec:errori}.  

In summary, the Monte Carlo simulations showed that the results from the AS09 scenario suggest a bias towards the overestimation of the mixing-length value, without the presence of any statistically significant trend, although a naive linear regression would indeed suggest a change in $\alpha_{\rm ml}$ of 0.23 for a 1 dex increase in [Fe/H]. 
The estimated mixing-length values are affected by a huge random uncertainty and are almost unconstrained in the explored range. 

Any attempt to calibrate the mixing-length parameter by means of $T_{\rm eff}$, [Fe/H], $\Delta \nu$, and $\nu_{\rm max}$ of field stars appears statistically unreliable, even in this very ideal case in which the artificial stars and the grid of models adopted for the recovery are in perfect agreement by construction. When dealing with real stars, even worse results should be expected.  

\subsection{Impact of the observational errors and evolutionary phase}\label{sec:errori}

This section explores the relevance of the observational uncertainties and the evolutionary phase on the recovered mixing-length values in the unbiased scenario AS09. 
To this purpose we repeated the $\alpha_{\rm ml}$ evaluation, assuming different uncertainties in the process described in Sect.~\ref{sec:as09}. 

First of all, we explored a scenario in which all the observational errors were over optimistically decreased to one-quarter of their nominal value. While this reduction in the asteroseismic constraints is certainly possible for selected sample of stars with exceptionally good quality (e.g. the LEGACY sample in \citealt{Lund2017}), it is far too optimistic for effective temperature and [Fe/H]. Indeed their errors where reduced to 12 K and 0.02 dex.

\begin{figure}
        \centering
        \includegraphics[height=8.4cm,angle=-90]{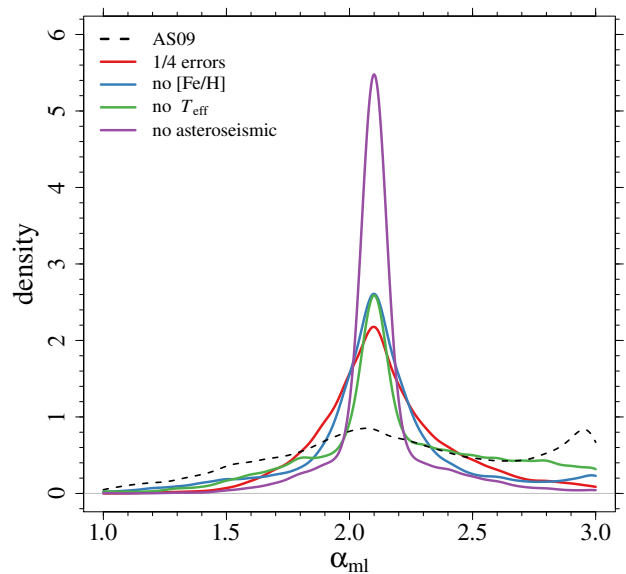}
        \caption{Density of the recovered $\alpha_{\rm ml}$ with different assumptions on the observational constraints. The dashed line refers to the standard reference scenario AS09. The red line considers simulations performed with errors reduced to one-quarter of their nominal value. The blue, green, and purple lines refer to recovery without errors on [Fe/H], effective temperature, and asteroseismic parameters, respectively.}
        \label{fig:effetto-errori}
\end{figure}

The results of this simulation are presented in Fig.~\ref{fig:effetto-errori}. The variance of the distribution is significantly reduced with respect to AS09 computations. The mean recovered $\alpha_{\rm ml}$ is $2.14 \pm 0.28$, therefore a reduction in the error to one-quarter leads to a reduction of the standard deviation to only about 50\% with respect to AS09 scenario.
The star-by-star variability is still notable. 

It seems therefore that the calibration of $\alpha_{\rm ml}$ from field stars  is still a questionable procedure, even assuming far overly optimistic observational errors and perfect agreement between the input physics and chemical composition of the mock data and stellar models.  

Then we studied various scenarios in which the errors in some observable constraints are neglected while keeping the others at their nominal value. Scenarios computed allowing individual errors in $T_{\rm eff}$, [Fe/H], and asteroseismic quantities produce very narrow and unbiased estimates of the mixing-length values and are not discussed further. 

When couples of observable constraints were allowed to vary within their errors, the best estimates were obtained keeping fixed the asteroseismic quantities. In fact the impossibility to bias the mass and radius of the stars -- which are mainly fixed by $\Delta \nu$ and $\nu_{\rm max}$ -- leads the algorithm to well recover the true value $\alpha_{\rm ml} = 2.1$.
The results when the errors in [Fe/H] and $T_{\rm eff}$ are neglected show somewhat inferior performances. In particular, in this last case the variability is large and has a tendency towards overestimation.

\begin{figure*}
        \centering
        \includegraphics[height=16.4cm,angle=-90]{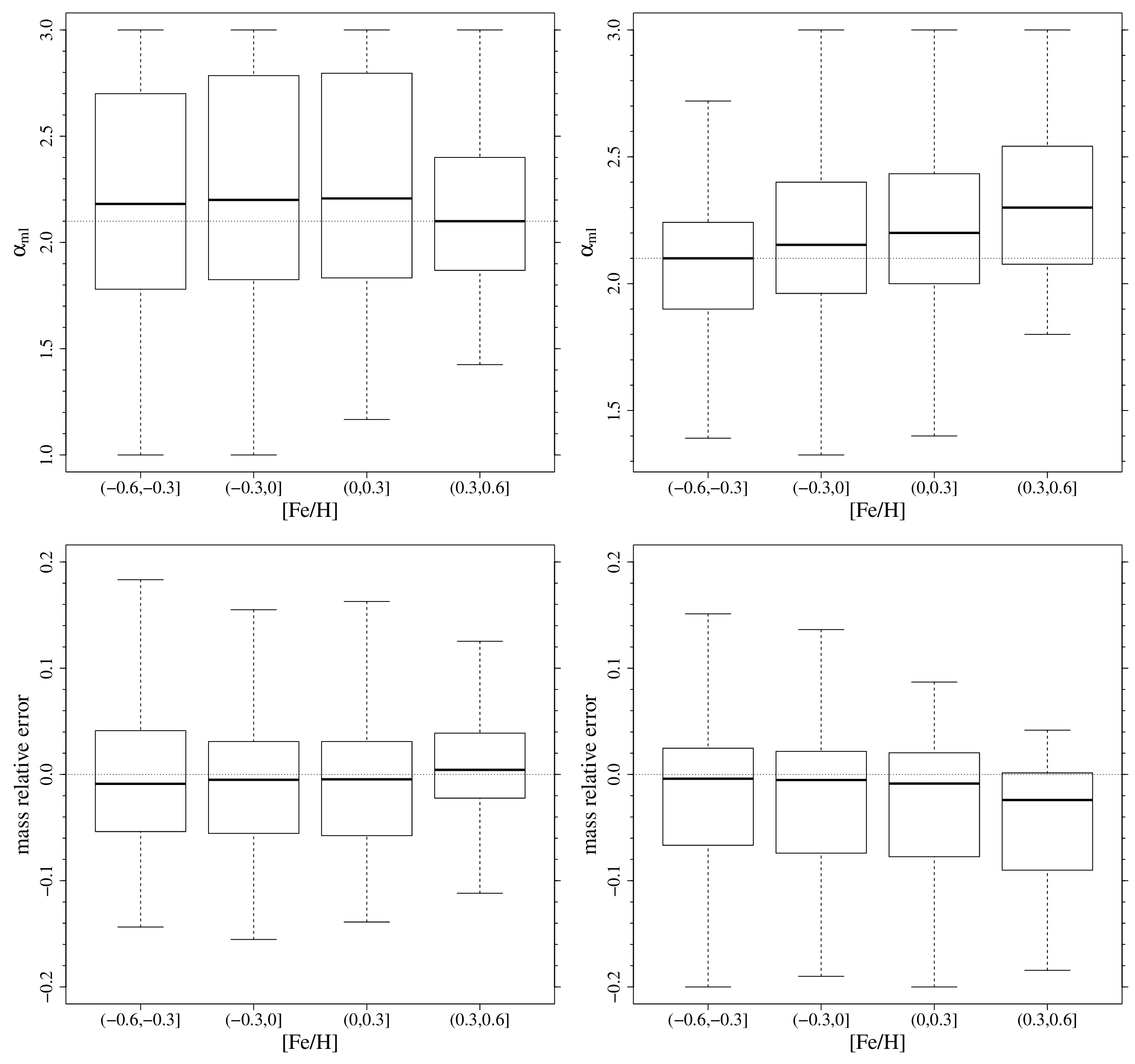}
        \caption{{\it Top row, left}: Boxplot of the estimated mixing-length values in the considered metallicity bin, only for MS stars. {\it Right}: As in the {\it left panel}, but for RGB models. {\it Bottom row, left}: Boxplot of the relative error on the reconstructed stellar mass in the considered metallicity bin, only for MS stars. {\it Right}: As in the {\it left panel}, but for RGB models.}
        \label{fig:effetto-fase}
\end{figure*}

The data from scenario AS09 in Sect.~\ref{sec:as09} were cumulatively analysed. However, some interesting differences in the $\alpha_{\rm ml}$ recovery arise in various evolutionary stages. We focus on the recovery of models in MS and in the RGB. 
Fig.~\ref{fig:effetto-fase} shows the recovery of the mixing-length values in the considered metallicity bins for different evolutionary phases. The left and right panels correspond to the MS and RGB models, respectively. 

\begin{figure}
        \centering
        \includegraphics[height=8.4cm,angle=-90]{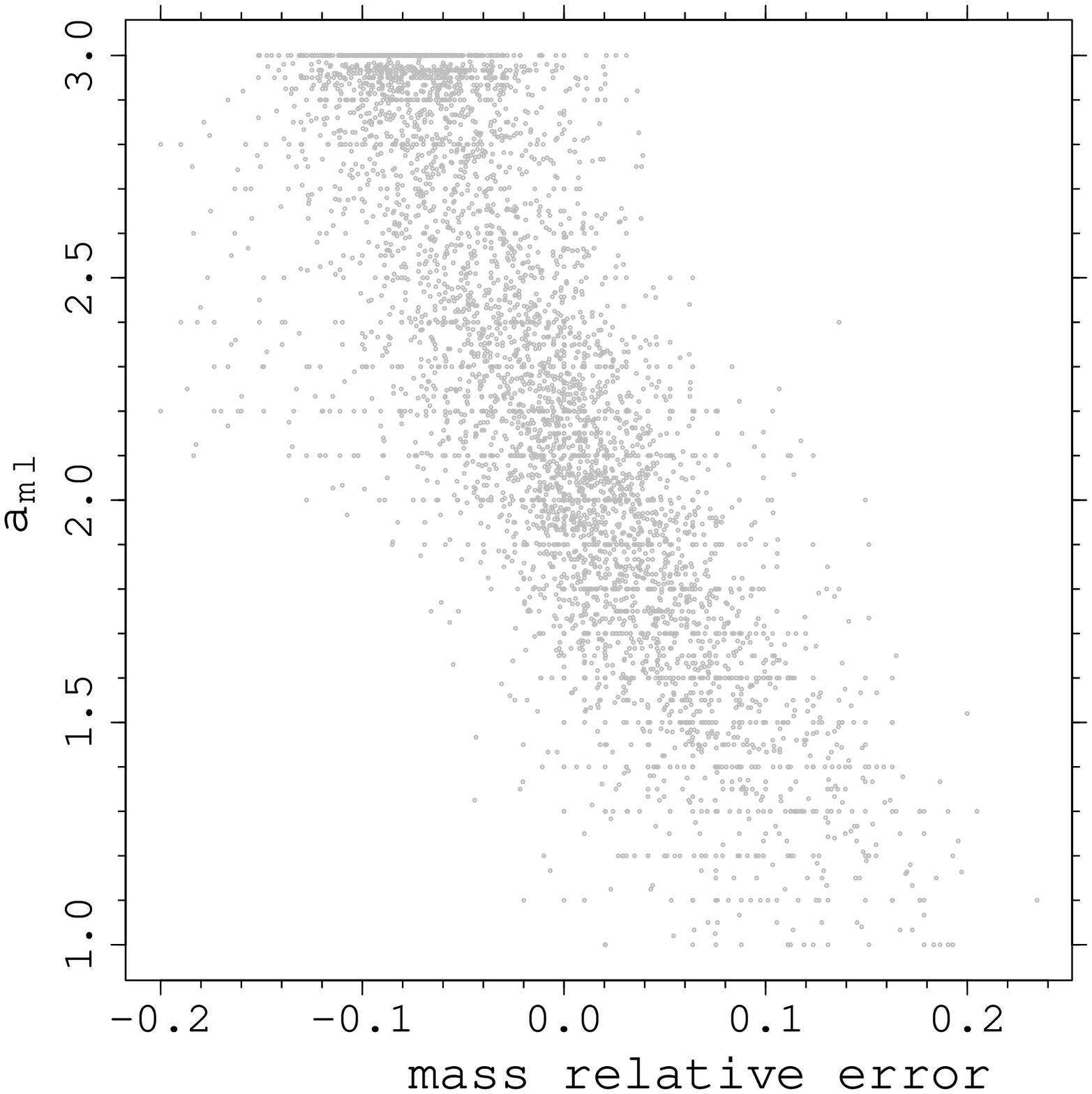}
        \caption{Scatterplot of the estimated $\alpha_{\rm ml}$ values as functions of the relative bias on the recovered stellar mass.}
        \label{fig:alpha-M}
\end{figure}

A fictitious trend appears in the RGB phase because of a known effect that occurs in the recovery of RGB stars \citep[see e.g.][]{bulge}. In this evolutionary phase the mass estimation is biased towards lower values due to non-homogeneous packing of the tracks in the HR diagram. The problem is more and more severe as the metallicity increases. This is shown in the bottom row in Fig.~\ref{fig:effetto-fase}, where the mass relative errors were reported for MS (left) and RGB (right). The strong dependence of the recovered $\alpha_{\rm ml}$ on the mass relative bias (shown in Fig.~\ref{fig:alpha-M}) explains this behaviour. A 1\% relative bias in the recovered stellar mass accounts for an $\alpha_{\rm ml}$  variation of about $-0.06$.

\subsection{GS98 scenario: heavy-element mixture mismatch}\label{sec:gs98}

In the GS98 scenario the sample of artificial stars was obtained from stellar models computed with the GS98 solar mixture and recovered on a grid of models adopting the AS09 solar mixture. Both datasets were computed at $\Delta Y/\Delta Z = 2.0$.
This allowed the exploration of a scenario in which the iron abundance [Fe/H] of data and grid are different for any given value of the global metallicity $Z$, thus introducing a distortion in the recovery. 
The heavy-element mixture in the Sun has been largely debated in recent years and various choices are currently adopted in the computation of stellar models. The results discussed in the following gives an initial idea of the impact of such an assumption on the mixing-length calibration.

The analysis of a similar scenario was considered particularly relevant in light of the results and comments in \citet{Salaris2018}. This work clearly showed that a change in the mixture adopted in the recovery can lead to biases in the results.  

\begin{figure*}
        \centering
        \includegraphics[height=16.0cm,angle=-90]{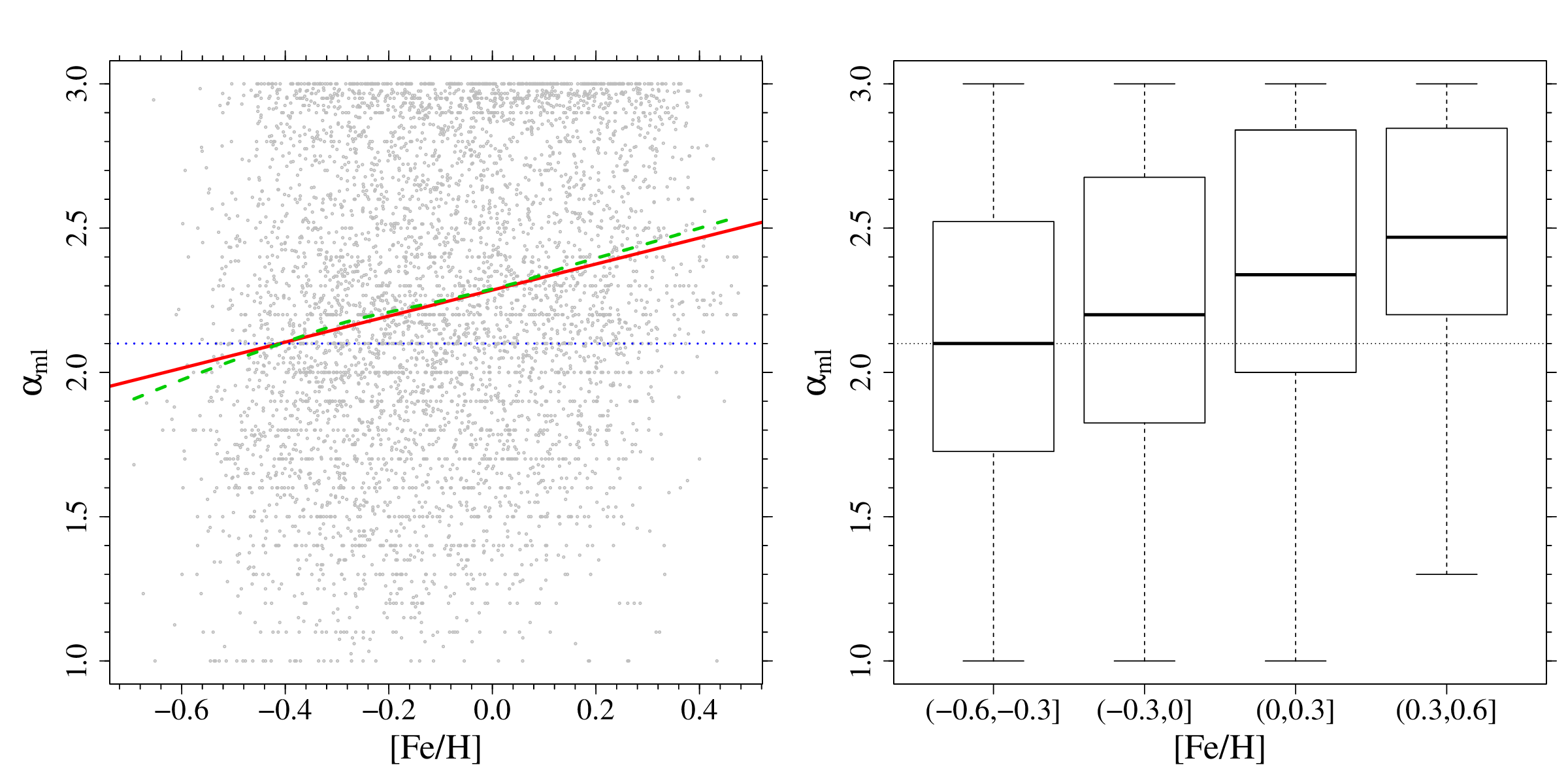}
        \caption{As in Fig.~\ref{fig:alpha-feh-as09}, but for GS98 scenario. }
        \label{fig:alpha-feh-gs98}
\end{figure*}

The results, presented in Fig.~\ref{fig:alpha-feh-gs98}, show the same huge spread discussed for the AS09 scenario, and a similar bias towards overestimation; the mean $\alpha_{\rm ml}$ is $2.24 \pm 0.54$  (about 9\% overestimation). The linear regression coefficient is $0.48 \pm 0.03$. An important difference with respect to the AS09 scenario is that the GS98 LOESS smoother closely matches the corresponding linear model. Most interestingly, the boxplot in the right panel of Fig.~\ref{fig:alpha-feh-gs98} shows a steady increase in the median value of the recovered $\alpha_{\rm ml}$ with [Fe/H] (see also Tab.~\ref{tab:med-alpha}). 
The estimated change in $\alpha_{\rm ml}$ in the explored metallicity range is about 0.37, resulting in a $\Delta \alpha_{\rm ml}$ of about 0.4 for a change of 1 dex in [Fe/H].

Therefore the trend with [Fe/H] of the estimated mixing-length parameter $\alpha_{\rm ml}$ in the GS98 scenario is statistically robust. However, it is completely spurious since the artificial stars have a constant $\alpha_{\rm ml} = 2.1$. Such an artefact in the recovered $\alpha_{\rm ml}$ is the direct consequence of the mismatch between the heavy-element mixture between artificial stars and grid of models. The detection of this artificial trend of the recovered mixing-length value versus the metallicity in a simple and controlled Monte Carlo set-up is a clear alarm bell that suggests great caution when attempting a calibration from real world objects. 
The direct determination of the abundances of the main elements is limited. As a consequence it is in practice infeasible to compute models with heavy-element mixtures that match those of all the objects of a large sample of field stars, thus exposing  biased results and artificial trends. 

\begin{figure*}
        \centering
        \includegraphics[height=16cm,angle=-90]{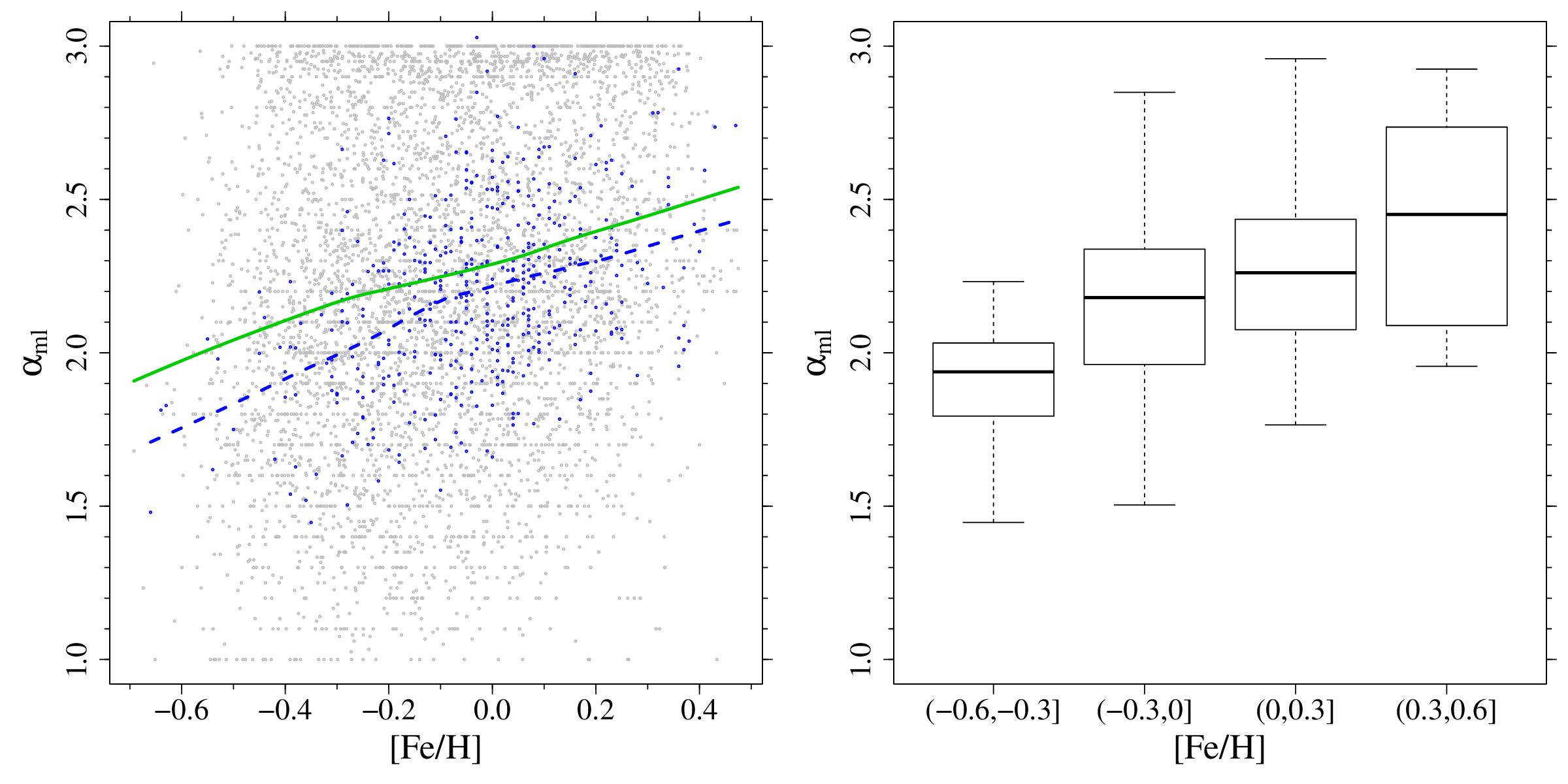}
        \caption{{\it Left}: Comparison of the $\alpha_{\rm ml}$ values obtained in the GS98 scenario (black dots) with those from \citet{Viani2018} (blue dots). The values from \citet{Viani2018} were offset by 0.4 to match the solar-scaled $\alpha_{\rm ml}$ of the GS98 scenario. The green solid and blue dashed lines are  the LOESS smoother for GS98 and \citet{Viani2018} datasets, respectively. {\it Right}: Boxplot of the \citet{Viani2018} results in the metallicity bins adopted in the present work.}
        \label{fig:cfr-basu}
\end{figure*}

Before further investigating the origin of the artificial trend it is worth noting that the trend detected in the GS98 scenario closely resembles that reported in \citet{Viani2018}.  Figure~\ref{fig:cfr-basu} shows the recovered mixing-length values from the GS98 scenario with superimposed the $\alpha_{\rm ml}$ derived for real field stars by \citet{Viani2018}. An offset of 0.4 was added to the latter to account for the differences in the solar calibrated $\alpha_{\rm ml}$ for the adopted stellar models. The LOESS smoothers in the figure show a very similar trend in the explored metallicity range. The right panel in the figure shows the \citet{Viani2018} results in the same metallicity bins adopted in this paper. These two trends were further subjected to statistical analysis that showed no significant difference between them (ANOVA $p$-value = 0.35).
The following sections are devoted to exploring the origin and robustness of the trend found from the Monte Carlo simulations. 

\section{Wrong assumptions in the $Y$ -- $Z$ -- [Fe/H] relation}\label{sec:feh}

In the previous section we showed that a spurious trend of the mixing-length value with the metallicity is induced in the GS98 scenario. In this case the artificial stars and grid of models used in the recovery adopt different heavy-element mixtures. The aim of the present section is to further investigate the reasons behind such a behaviour. 

A mismatch between the heavy-element mixture of the stars and that adopted to compute the grid of models used to calibrate the mixing-length parameter has several effects. A change of the heavy-element mixture affects the mean Rosseland opacity at a given metallicity, which in turn modifies the evolutionary timescale and stellar track location in the HR diagram and in the observables hyperspace. Moreover, if the CNO-element sum is modified, it also impacts the hydrogen-burning efficiency \citep[see e.g.][]{Salaris1993, VandenBerg2012}. 

A second effect is a modification of the three-way relation $Y$ -- $Z$ -- [Fe/H]. Therefore the [Fe/H] scale is altered, so that different $Z$ values correspond to the same [Fe/H] and vice versa. Similarly, different initial helium abundances $Y$ correspond to the same [Fe/H] because of the assumed linear relation in Eq.~(\ref{eq:dydz}). These modifications have a direct impact on the stellar model effective temperatures. Given the strong dependence of this last quantity on the mixing-length value, it is clear that an effect that leads to different temperature biases as functions of the metallicity in turn directly translates into a differential effect in the recovered mixing-length values.

In other words, the adoption of a wrong heavy-element mixture in the computation of stellar models adopted for the recovery induces a systematic discrepancy between the models and synthetic stars which is a function of the metallicity. As a consequence, in a maximum-likelihood fitting procedure -- regardless the details of the implemented algorithm -- such a discrepancy is counterbalanced by a change in the mixing-length parameter of the best-fitted model which, in turn, depends on metallicity. The result is a spurious metallicity trend in the calibrated mixing-length parameter $\alpha_{\rm ml}$.

While the relevance of the modification of the radiative opacities is hard to explore given the multiple changes in the element abundance from one mixture to another, the modification of the $Y$ -- $Z$ -- [Fe/H] relation is much simpler to investigate. The intertwined dependence of the three variables makes the disentanglement of their effects problematic, because of a modification of one of these impacts on all the others, as results from Eqs.~(\ref{eq:dydz}) and (\ref{eq:Z}). To shed some light in the expected behaviours due to possible modification of the three-way relation, the following sections focus individually on the effects caused by a change in $\Delta Y/\Delta Z$ (Sect.~\ref{sec:bias-dydz}) and by the modification of the [Fe/H] -- $Z$ relation (Sect.~\ref{sec:feh-scale}) in the grid of models used in the recovery.

\subsection{Effect of a biased $\Delta Y/\Delta Z$ in the recovery grid}\label{sec:bias-dydz}

A first effect that is worth exploring is the difference in the initial helium abundance at a given [Fe/H] due to a different heavy-element mixture. As well known, the initial helium abundance plays a key role in the evolution of stars. Keeping fixed all the other inputs, a change of the initial helium abundance significantly affects the location of stellar tracks in the hyperspace of the observables (in our case $T_{\rm eff}$, $\Delta \nu$, $\nu_{\rm max}$, [Fe/H]). In some evolutionary phases, such a displacement mimics the effect of a change of the mixing-length parameter. Thus, a wrong assumption of the initial helium abundance necessarily affects the calibration of the mixing-length parameter from field stars.  

\begin{figure}
        \centering
        \includegraphics[height=8.2cm,angle=-90]{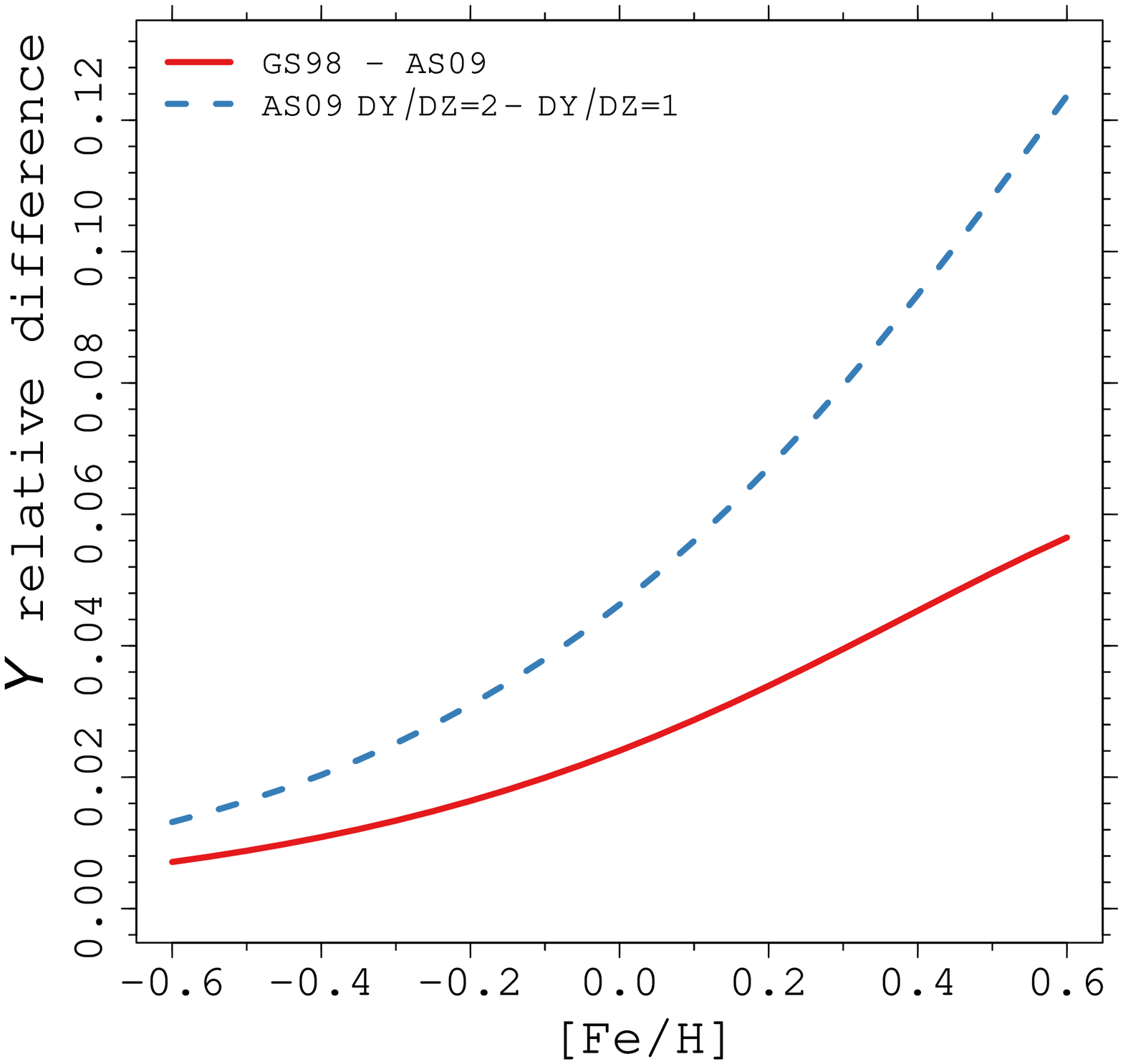}
        \caption{Relative differences in the initial helium abundances (assuming the same $\Delta Y/\Delta Z = 2.0$) between the AS09 and GS98 mixtures (red solid line) and between  $\Delta Y/\Delta Z = 2.0$ and 1.0, assuming the same mixture AS09 (blue dashed line).  }
        \label{fig:Y-reldiff}
\end{figure}

Figure~\ref{fig:Y-reldiff} shows the relative difference in the initial helium abundances between AS09 and GS98 solar mixtures as a function of metallicity [Fe/H], assuming the same helium-to-metal enrichment ratio $\Delta Y/\Delta Z = 2.0$. While the relative difference is lower than 1\% for [Fe/H] $\lesssim -0.4$,  it is above 5\% for the highest considered metallicities. This clearly induces a bias in the calibration process of the mixing-length for the GS98 scenario, which depends on the metallicity [Fe/H].

\begin{figure*}
        \centering
        \includegraphics[height=16.4cm,angle=-90]{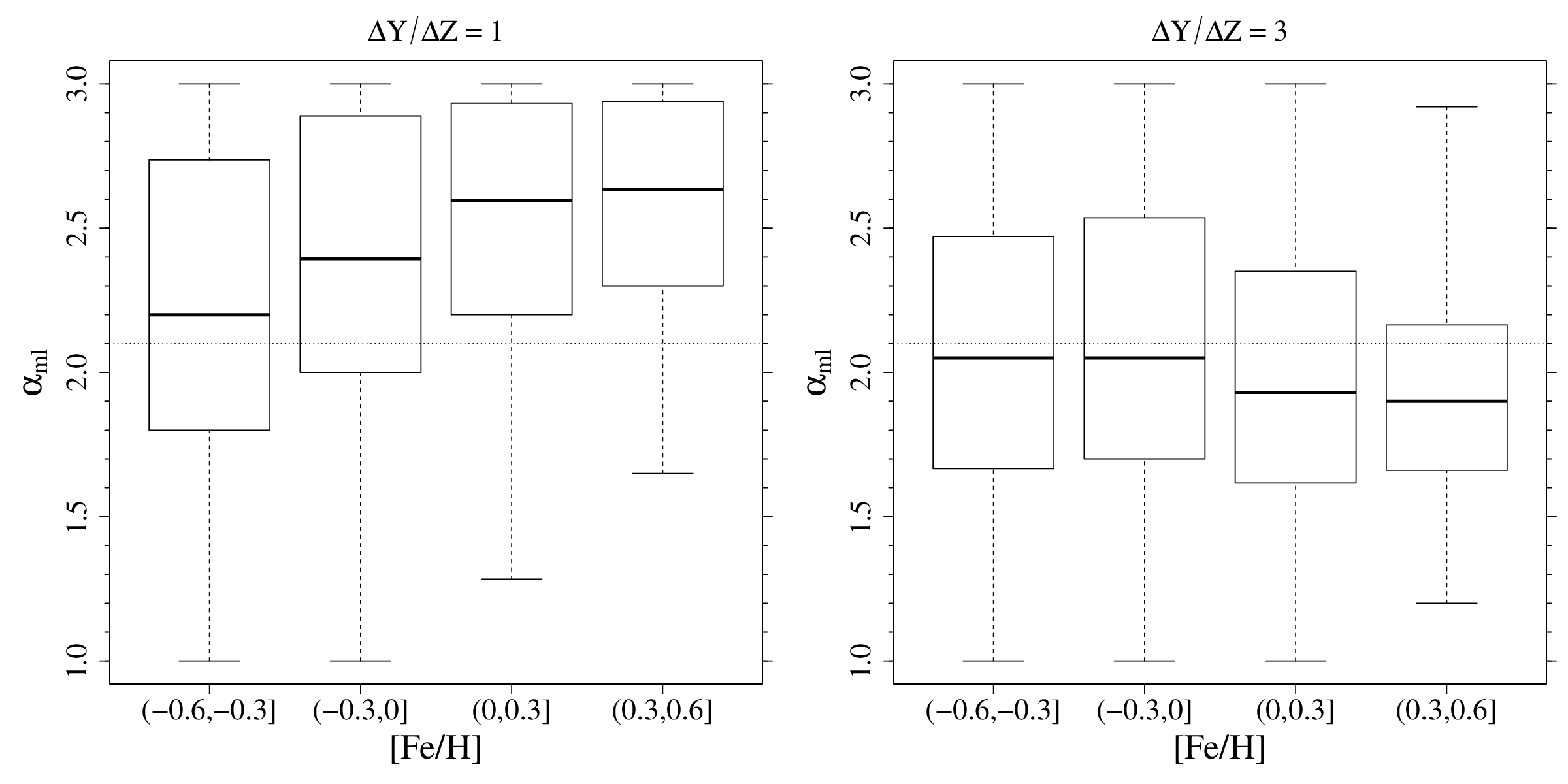}\\
        \includegraphics[height=16.4cm,angle=-90]{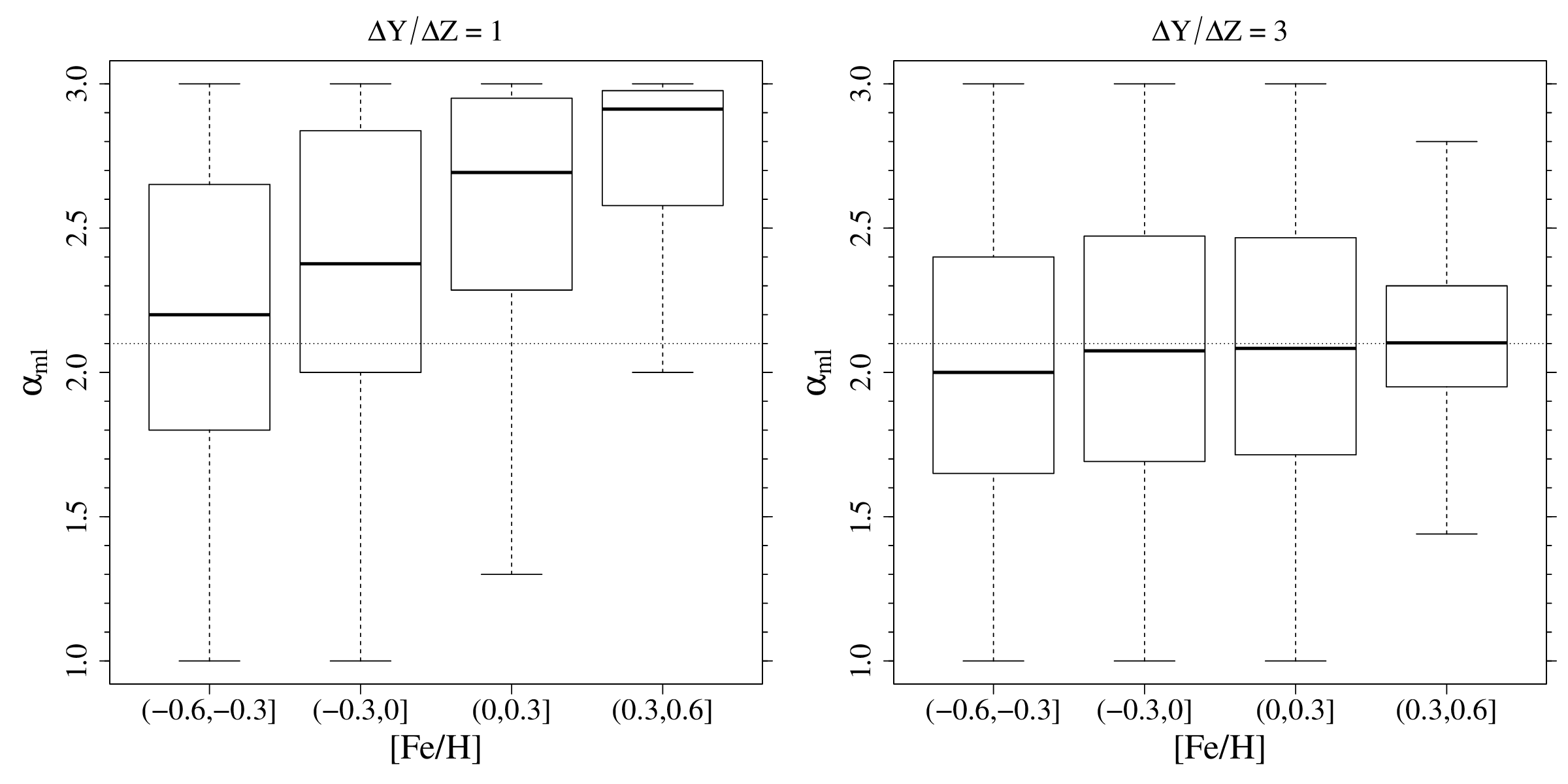}
        \caption{{\it Top row, left}: Boxplot of the estimated $\alpha_{\rm ml}$ values from AS09 scenario adopting in the recovery a grid with models at $\Delta Y/\Delta Z = 1.0$. {\it Right}: Same as in the left panel, but recovering with models at $\Delta Y/\Delta Z = 3.0$. In all the cases artificial stars are sampled from a grid with $\Delta Y/\Delta Z = 2.0$. {\it Bottom row}: Same as in the  top row, but for the GS98 scenario.}
        \label{fig:effettoDYDZ}
\end{figure*}

In the light of this difference  
it is also interesting to explore in some detail the influence of wrong assumptions on the helium-to-metal enrichment ratio $\Delta Y/\Delta Z$ -- at fixed mixture -- in the estimation process. Such a quantity is in fact quite uncertain because the commonly adopted value is 2 $\pm$ 1 \citep{pagel98,jimenez03,flynn04,gennaro10}  but several authors claim even larger errors. 

A change of the adopted $\Delta Y/\Delta Z$ at fixed heavy-element mixture causes a variation of the initial helium abundance at a given [Fe/H] qualitatively similar to that previously shown caused by a change of the heavy-element mixture at fixed $\Delta Y/\Delta Z$. As an example Fig.~\ref{fig:Y-reldiff} shows the effect of decreasing the $\Delta Y/\Delta Z$ value from 2 to 1, taking fixed the mixture AS09. An approximative constant scaling factor of two exists between the curves in the figure, therefore this change qualitatively mimics the differences between the GS98 and AS09 scenarios. 

A systematic discrepancy  between the initial helium abundance adopted to compute the stellar models used in the fitting procedure and that actually present in the synthetic stars necessarily biases the fitting results. Such a discrepancy alters the location of stellar tracks in the hyperspace of the observables with respect to the "real" locations. The fitting algorithm is thus forced to compensate these displacements by biasing the estimated masses and $\alpha_{\rm ml}$ values. 

To quantify the relevance of the current uncertainty affecting the initial helium abundance on the mixing-length estimate, we analysed four non-standard scenarios: two for the AS09 heavy-element mixture and two for the GS98 mixture.

For the AS09 mixture we used the same mock dataset described in Sect.\ref{sec:as09} made by artificial stars computed with the AS09 heavy-element mixture, mixing-length  $\alpha_{\rm ml}$ = 2.1 and $\Delta Y/\Delta Z$ = 2.0. In the fitting procedure we then applied the SCEPtER pipeline to this dataset but relying on a non-standard grid of models. In the AS09-Y1 scenario we used models computed with the same mixture but $\Delta Y/\Delta Z = 1.0$, while we adopted $\Delta Y/\Delta Z = 3.0$ for the AS09-Y3 scenario. These two cases allow us to explore the impact on the mixing-length calibration of adopting the "true" heavy-element mixture but the wrong -- although within the current uncertainty -- $\Delta Y/\Delta Z$. 

We also performed the same exercise for the GS98 mixture, for which the mock dataset is the same as that described in Sect.\ref{sec:gs98} made by artificial stars computed with the GS98 heavy-element mixture, mixing-length  $\alpha_{\rm ml}$ = 2.1, and $\Delta Y/\Delta Z$ = 2.0. In the GS98-Y1 scenario the grid of models used in the SCEPtER pipeline adopts the AS09 heavy-element mixture  but $\Delta Y/\Delta Z = 1.0$, while in the GS98-Y3 scenario $\Delta Y/\Delta Z = 3.0$ was assumed.

\begin{figure*}
        \centering
        \includegraphics[height=16.4cm,angle=-90]{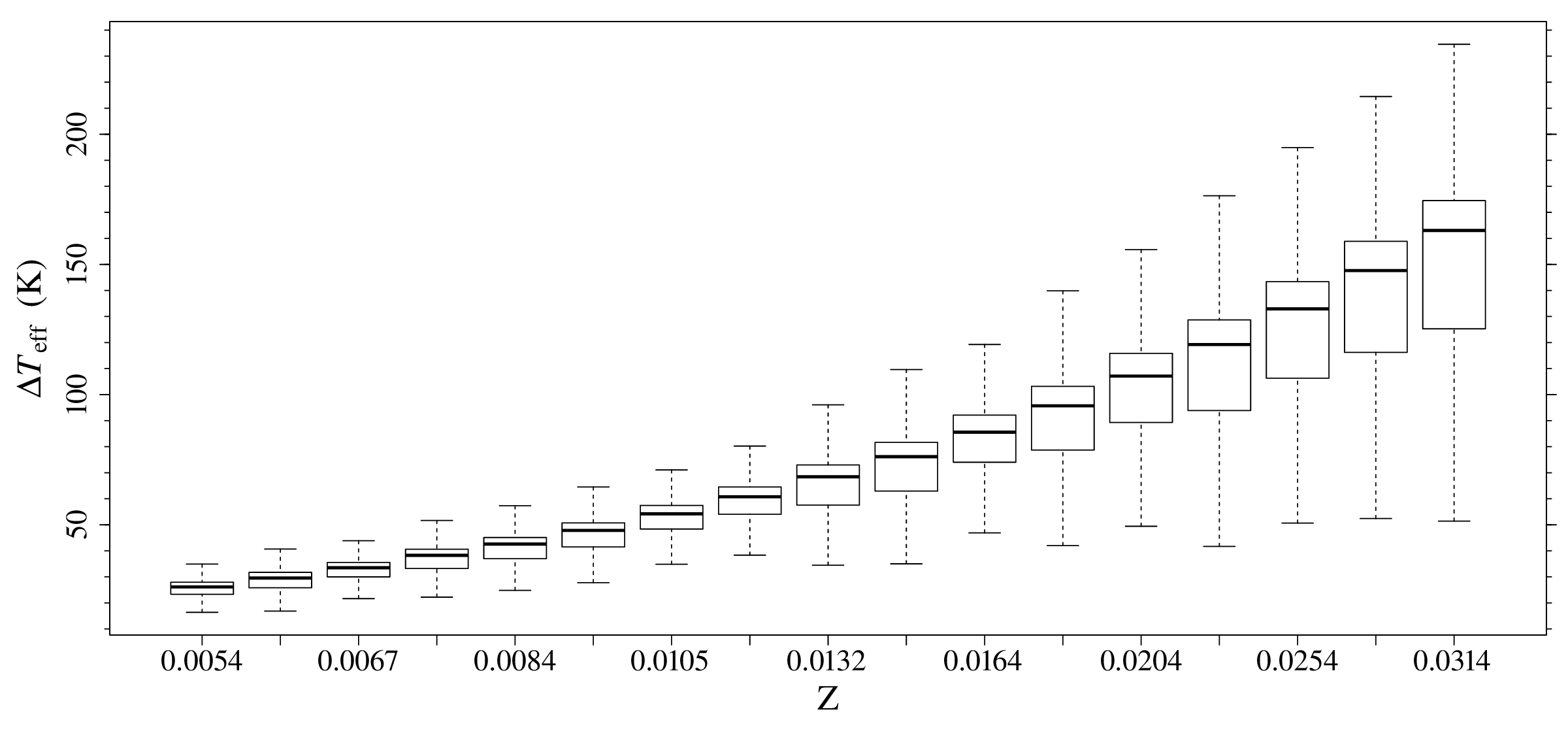}
        \caption{Difference in the effective temperature of homologous stellar models between grids at $\Delta Y/\Delta Z = 2.0$ and $\Delta Y/\Delta Z = 1.0$, for various values of the metallicity $Z$.}
        \label{fig:dydz2-offset-Teff}
\end{figure*}

The resulting estimated mixing-length values are presented in Fig.~\ref{fig:effettoDYDZ} and Tab.~\ref{tab:med-alpha}. Results from scenario AS09-Y1 show that the underestimation of the initial helium abundance leads to an overestimation of the mean mixing-length value at a given [Fe/H] and to infer a spurious trend of $\alpha_{\rm ml}$  with [Fe/H]. Such a result can be easily understood because at fixed mass and evolutionary stage, as the initial helium abundance decreases the same occurs for the effective temperature. This in turn leads to a concurrent overestimation of the $\alpha_{\rm ml}$ value to compensate. Moreover, the change in the $\Delta Y/\Delta Z$ value causes a relative difference in the initial helium abundances at fixed [Fe/H] as shown in Fig.~\ref{fig:Y-reldiff}. Therefore the compensation effect in $\alpha_{\rm ml}$ must increase with [Fe/H]. As an example of this effect, Fig.~\ref{fig:dydz2-offset-Teff} shows the difference in the effective temperature between homologous models with $\Delta Y/\Delta Z = 2.0$ and $\Delta Y/\Delta Z = 1.0$, as a function of the initial metallicity $Z$. The difference is clearly not constant and increases at higher metallicity, therefore requiring larger $\Delta \alpha_{\rm ml}$ to compensate for this shift.  

It is therefore understandable that an underestimation of $\Delta Y/\Delta Z$  induces the trend shown in the left panel in the top row of Fig.~\ref{fig:effettoDYDZ}, similar to that for the GS98 scenario.  For scenario AS09-Y1, the detected variation of $\alpha_{\rm ml}$  for a change of 1 dex in [Fe/H] is 0.48. This value is probably affected by edge effects in the highest metallicity bin because many $\alpha_{\rm ml}$ estimates cluster at the upper value (i.e. 3.0) allowed in the fitting process.

The results from the scenario AS09-Y3 (right panel, top row in Fig.~\ref{fig:effettoDYDZ}) show an opposite behaviour, which has a tendency towards mixing-length underestimation and a negative trend of $\alpha_{\rm ml}$  with the metallicity. 

The same qualitative behaviours were detected for GS98-Y1 and GS98-Y3 scenarios. The left panel in the bottom row of Fig.~\ref{fig:effettoDYDZ} shows an even steeper trend of the recovered $\alpha_{\rm ml}$ with the metallicity, while this trend cancels out in the GS98-Y3 scenario (right panel, bottom row in Fig.~\ref{fig:effettoDYDZ}). In this last case the effects of the change in the heavy-element mixture and the mismatch in the  initial helium abundance specification nearly compensate each other.
 
In summary, the results described in this section show that the current uncertainty in the helium-to-metal enrichment ratio $\Delta Y/\Delta Z$ is large enough to induce a spurious trend in the estimated mixing-length value  $\alpha_{\rm ml}$  with [Fe/H], even in this very ideal case where the true $\alpha_{\rm ml}$  is fixed and all the other inputs/parameters adopted in stellar models are in perfect agreement with the artificial stars.

\subsection{Effect of a rigid shift in the [Fe/H] -- $Z$ relation}\label{sec:feh-scale}

\begin{figure*}
        \centering
        \includegraphics[height=16.6cm,angle=-90]{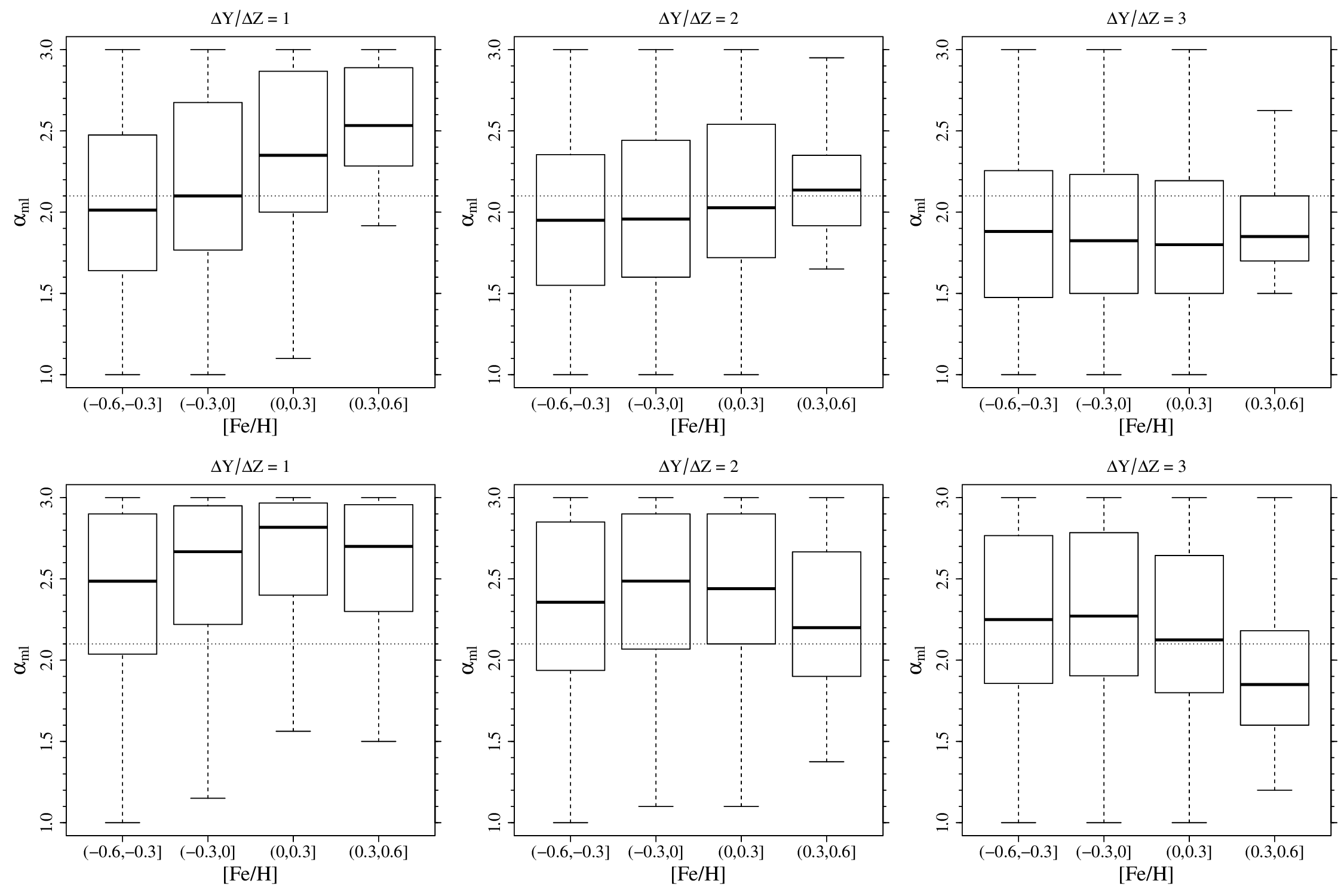}
        \caption{{\it Top row,  left}:  Boxplot of the estimated mixing-length values for an artificial offset of $-0.1$ dex in the observed [Fe/H] values (scenario AS09-Mm) in the adopted metallicity bins. The recovery was performed with a grid computed at $\Delta Y/\Delta Z = 1.0$. {\it Middle}: Same as in the left panel, but with a recovery grid computed with $\Delta Y/\Delta Z = 2.0$. {\it Right}: Same as in the left panel, but with a recovery grid computed with $\Delta Y/\Delta Z = 3.0$
        {\it Bottom row}: Same as in the top row, but for an artificial offset of $+0.1$ dex in the [Fe/H] (AS09-Mp scenario). }
        \label{fig:offsetFeH}
\end{figure*}

A different way to look at the same question concerning the variation in the $Y$ -- $Z$ -- [Fe/H] relation due to heavy-element mixture mismatch between observed stars and recovery grid 
is to focus directly on the [Fe/H] to global metallicity $Z$ relation.

Fixing $\Delta Y/\Delta Z = 2.0$ and [$\alpha$/Fe] = 0.0, the same value of [Fe/H] = 0.0 corresponds to $Z$ = 0.013 and 0.016 ($Y$ = 0.27 and 0.28) for solar mixtures AS09 and GS98,  respectively. Analogously, the values $Z = 0.013$, $Y = 0.274$ that provide [Fe/H] = 0.0 for AS09 solar mixture give [Fe/H] = $-0.1$ when adopting the GS98 solar mixture. 

A direct way to explore the impact of a similar constant shift in the AS09 scenario is to modify artificially the [Fe/H] values of the mock data. We tested two different constant shifts, namely by $+0.1$ dex (scenario AS09-Mp) and by $-0.1$ dex (scenario AS09-Mm). We then applied to these new mock datasets the SCEPtER pipeline relying on the standard grid of models computed with the AS09 heavy-element mixture  and $\Delta Y/\Delta Z = 2.0$. Additional analyses adopting grids with $\Delta Y/\Delta Z = 1.0$ and 3.0 were also explored.

The top row of Fig.~\ref{fig:offsetFeH} shows the reconstruction of an artificial perturbation of the mock data [Fe/H] by $-0.1$ dex, while the bottom row reports the equivalent results from a perturbation of $+0.1$ dex. It is interesting to note that the middle panel in the top row, corresponding to the recovery at $\Delta Y/\Delta Z = 2.0,$ shows a steady trend of $\alpha_{\rm ml}$ with [Fe/H], which confirms the statement that a whatever modification of the $Y$ -- $Z$ -- [Fe/H] relations can potentially induce an artificial trend in the recovered mixing-length values with [Fe/H].     

The mean $\alpha_{\rm ml}$ value from the AS09-Mm at $\Delta Y/\Delta Z = 2.0$ is 2.00 (standard deviation 0.53), showing an expected underestimation with respect to the reference value 2.10. In fact, the $-0.1$ dex shift in [Fe/H] of the mock data forces the fitting algorithm to explore lower metallicity tracks with respect to the standard scenario from which mock data are sampled. As a consequence the effective temperatures of the recovery models are too high with respect to the data. This discrepancy leads to adjusting the mixing-length towards lower values to match the effective temperature and the asteroseismic parameters of the artificial stars. 

An opposite behaviour occurs in the AS09-Mp scenario because the median recovered mixing-length value is 2.38 (standard deviation 0.48). 
A rough estimate of  the variation of mixing-length with [Fe/H] in scenario AS09-Mm can be obtained considering the differences in the $\alpha_{\rm ml}$ median values of the two extreme considered metallicity bins (with $\Delta Y/\Delta Z = 2.0$). The resulting estimated trend for a variation of 1 dex in [Fe/H] is about 0.21, which is one-half of that from GS98 scenario. Therefore, it seems that the simple calibration of the [Fe/H] scale plays a relevant role in the trend development.

The exercises performed in this section and in Sect.~\ref{sec:bias-dydz} have the merit of highlighting that even a simple offset in the observed metallicity with respect to the scale adopted in the grid of models used in the recovery can potentially induce artificial trends with [Fe/H] in the estimated mixing-length. Even if the models and data match for every other physics and chemical input, this mismatch in the metallicity forces the estimation algorithm 
to explore a wrong portion in the grid hyperspace thus providing biased estimates of the stellar masses and mixing-length values. 

In summary, a change in the $Y$ -- $Z$ -- [Fe/H] relation caused by a rigid shift in the observed [Fe/H] values or by a change in the $\Delta Y/\Delta Z$ value has the potential to induce a spurious trend on the recovered mixing-length parameter with [Fe/H]. 

In light of these results the procedure of mixing-length calibration from field stars, at least with the adopted observational constraints ($T_{\rm eff}$, $\Delta \nu$, $\nu_{\rm max}$, [Fe/H]), is not theoretically grounded even in the ideal configuration, where all the other input physics/parameters adopted in stellar models are in perfect agreement with the stars.   

\section{Other potential spurious trend makers}\label{sec:other-trends}

In spite of the great improvement in accuracy and precision of stellar models, they are still affected by non-negligible uncertainties. As a consequence, when real stars are analysed rather than artificial stars, systematic discrepancies between the theoretical models and data should be expected. On the other hand, the results described in Sect. \ref{sec:feh} proved that a discrepancy between the $Y$ -- $Z$ -- [Fe/H] relation adopted to compute the models and that actually present in stars leads to a statistically significant trend with metallicity in the inferred mixing-length parameter which is nevertheless totally spurious, i.e. a mere artefact. 

It is therefore plausible that similar spurious trends might be induced by other systematic differences between real stars and stellar models. Any systematic error that causes a displacement of stellar tracks in the hyperspace of the observables whose extent depends on [Fe/H] leads to artefacts in the inferred mixing-length parameter $\alpha_{\rm ml}$, spurious metallicity dependence included. In the following subsections, we analysed in particular the effects of changing the efficiency of microscopic diffusion, artificially altering the effective temperature scale, and modifying the outer boundary conditions adopted to compute the grid of stellar models.  

\subsection{Effect of microscopic diffusion}\label{sec:diffusion}

An example of a potential source of systematic discrepancy is the efficiency of microscopic diffusion. Different treatments of diffusion in stellar computations result in different $T_{\rm eff}$ and surface [Fe/H] at a given mass, age, and initial chemical composition ($Y$ and $Z$), thus altering the [Fe/H] to global $Z$ relation. At variance with the AS09-Mm and AS09-Mp scenarios discussed in Sect.~\ref{sec:feh-scale}, the discrepancy between the true [Fe/H] actually present in real stars and those predicted in the grid of models is not constant over the whole grid but it varies with the evolutionary phase, mass, and initial chemical composition. 

\begin{figure*}
        \centering
        \includegraphics[height=16.4cm,angle=-90]{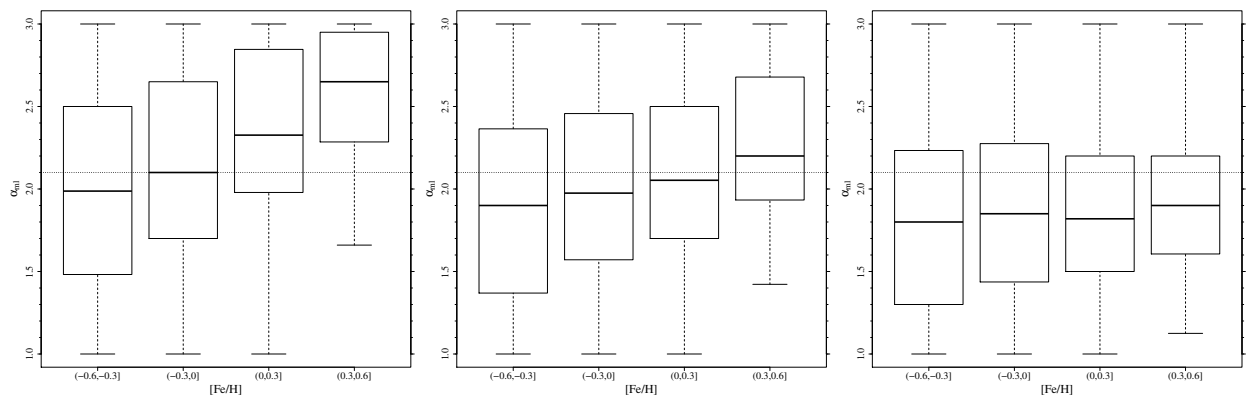}
        \caption{{\it Left}: Boxplot of the $\alpha_{\rm ml}$ estimated by means of a grid of models identical to AS09 but with no change on the surface [Fe/H] with respect to the ZAMS value. The adopted $\Delta Y/\Delta Z$ is 1.0. {\it Middle}: Same as in the left panel, but adopting $\Delta Y/\Delta Z = 2.0$ in the recovery. {\it Right}: Same as in the left panel, but adopting $\Delta Y/\Delta Z = 3.0$ in the recovery.}
        \label{fig:effetto-diffusione}
\end{figure*}

To quantify the relevance of microscopic diffusion on the mixing-length estimate, we analysed an additional non-standard scenario where the artificial stars took into account diffusion while the grid of models neglected it. In greater detail, we used the same mock dataset of AS09 scenario described in Sect.\ref{sec:as09} computed with $\alpha_{\rm ml} = 2.1$, $\Delta Y/\Delta Z = 2.0$ and took into account microscopic diffusion. We then applied to this dataset the SCEPtER pipeline but relied on a grid of models that simulate the neglecting of diffusion. 
To avoid the heavy computational burden of calculating a full set of stellar models which neglect microscopic diffusion, we exploited the results by \citet{scepter1}, which showed that about two-thirds of the distortions coming from totally neglecting microscopic diffusion can be simulated by neglecting the variation of the surface [Fe/H] during the evolution. This choice does not account for the change in the $T_{\rm eff}$ and evolutionary timescale induced by the occurrence of atomic diffusion.
Therefore we adopted in the recovery the same AS09 grid, but with [Fe/H] values kept constant at their zero age main sequence (ZAMS) value (scenario AS09-nd). The fitting procedure was performed for three different choices of the helium-to-metal enrichment ratio adopted in the grid of models, namely $\Delta Y/\Delta Z = 1.0$, 2.0, and 3.0. 

The middle panel in Fig.~\ref{fig:effetto-diffusione} shows that neglecting diffusion induces a spurious trend with [Fe/H] in the inferred mixing-length parameter $\alpha_{\rm ml}$ even when the same unbiased $\Delta Y/\Delta Z = 2.0$ is adopted in the grid of models as in the data. An increase of 1 dex in [Fe/H] roughly corresponds to a 0.3 increase in $\alpha_{\rm ml}$.

\subsection{Effect of a systematic error in the $T_{\rm eff}$ scale}\label{sec:teff}

It is well known that the effective temperature of stars, which is one of the observational constraints adopted in our analysis, is subject to a non-negligible systematic uncertainty.
Although it is not uncommon to find quoted errors in the effective temperature of the order of some tens of degrees, typically a comparison of results by different authors shows discrepancies even larger than 100 K \citep[see e.g.][]{Ramirez2005,Masana2006,Casagrande2010, Schmidt2016}. 
 
To explore the effect of a possible mismatch in the $T_{\rm eff}$ scale, we analysed two additional non-standard scenarios. We used the same mock dataset of AS09 scenario as described in Sect.~\ref{sec:as09}, but perturbing the effective temperatures. More in detail, we studied the case of a shift of $+100$ K in the AS09-Tp scenario and of $-100$ K in the AS09-Tm one. 
We then applied to these datasets the SCEPtER pipeline, relying on the standard grids of models.  

\begin{figure*}
        \centering
        \includegraphics[height=16.4cm,angle=-90]{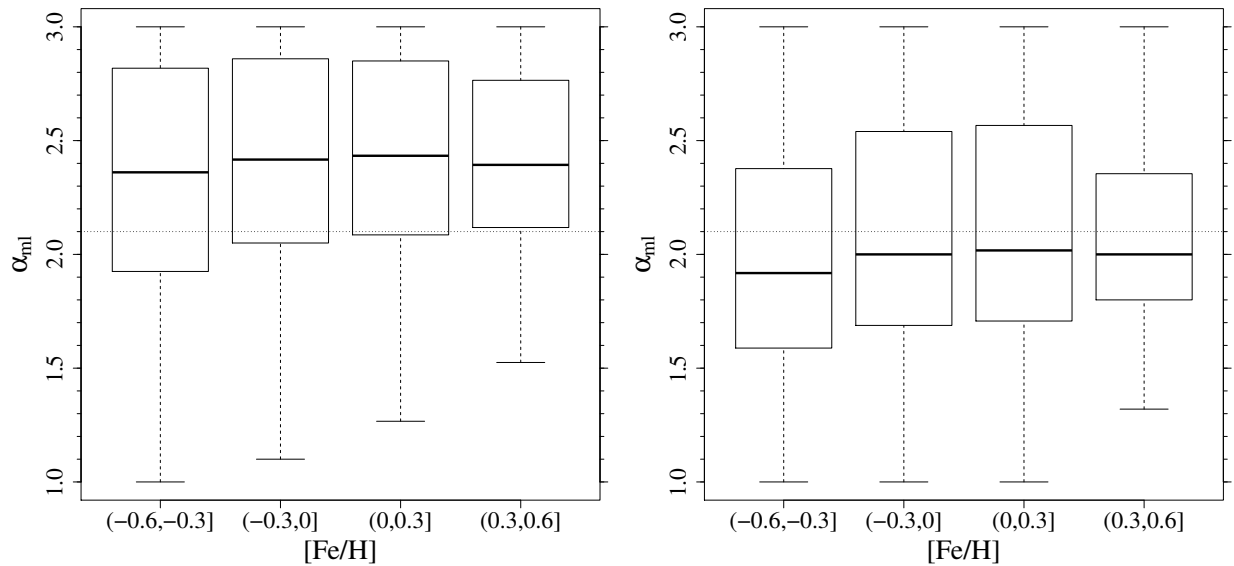}
        \caption{{\it Left}: Boxplot of the estimated $\alpha_{\rm ml}$ for an offset in the synthetic effective temperatures of $+100$ K. {\it Right}: Same as in the left panel, but for an offset of $-100$ K.       }
        \label{fig:effetto-teff}
\end{figure*}

Figure~\ref{fig:effetto-teff} shows that an offset in the effective temperature scale causes a systematic bias in the recovered mixing-length parameter. However, it does not induce a statistically significant trend with [Fe/H]. The reason is that in this scenario the displacement of stellar tracks is constant regardless of the metallicity and thus a constant offset in $\alpha_{\rm ml}$ is sufficient to compensate the bias. The left panel of Fig.~\ref{fig:effetto-teff} shows that an offset of $+100$ K (scenario AS09-Tp) leads to an overestimation of $\alpha_{\rm ml}$ with mean value of about 0.27 (the true value being $\alpha_{\rm ml} = 2.1$). The right panel of Fig.~\ref{fig:effetto-teff} shows that an offset of $-100$ K (scenario AS09-Tm) causes an underestimation of $\alpha_{\rm ml}$ of about $-0.1$.

Although the rigid offset in the effective temperature can be compensated by a rigid shift in the recovered mixing-length values, a different behaviour is expected in the presence of a trend in $T_{\rm eff}$ with [Fe/H]. In this case the mixing-length values develop an equivalent trend to compensate the differential shifts. Such a scenario occurs often when comparing different estimates of atmospheric parameters between different surveys. As an example  \citet{Anguiano2018} found a clear trend of $T_{\rm eff}$ with [Fe/H] comparing the results between the APOGEE and LAMOST surveys.

\begin{figure*}
        \centering
        \includegraphics[height=16.4cm,angle=-90]{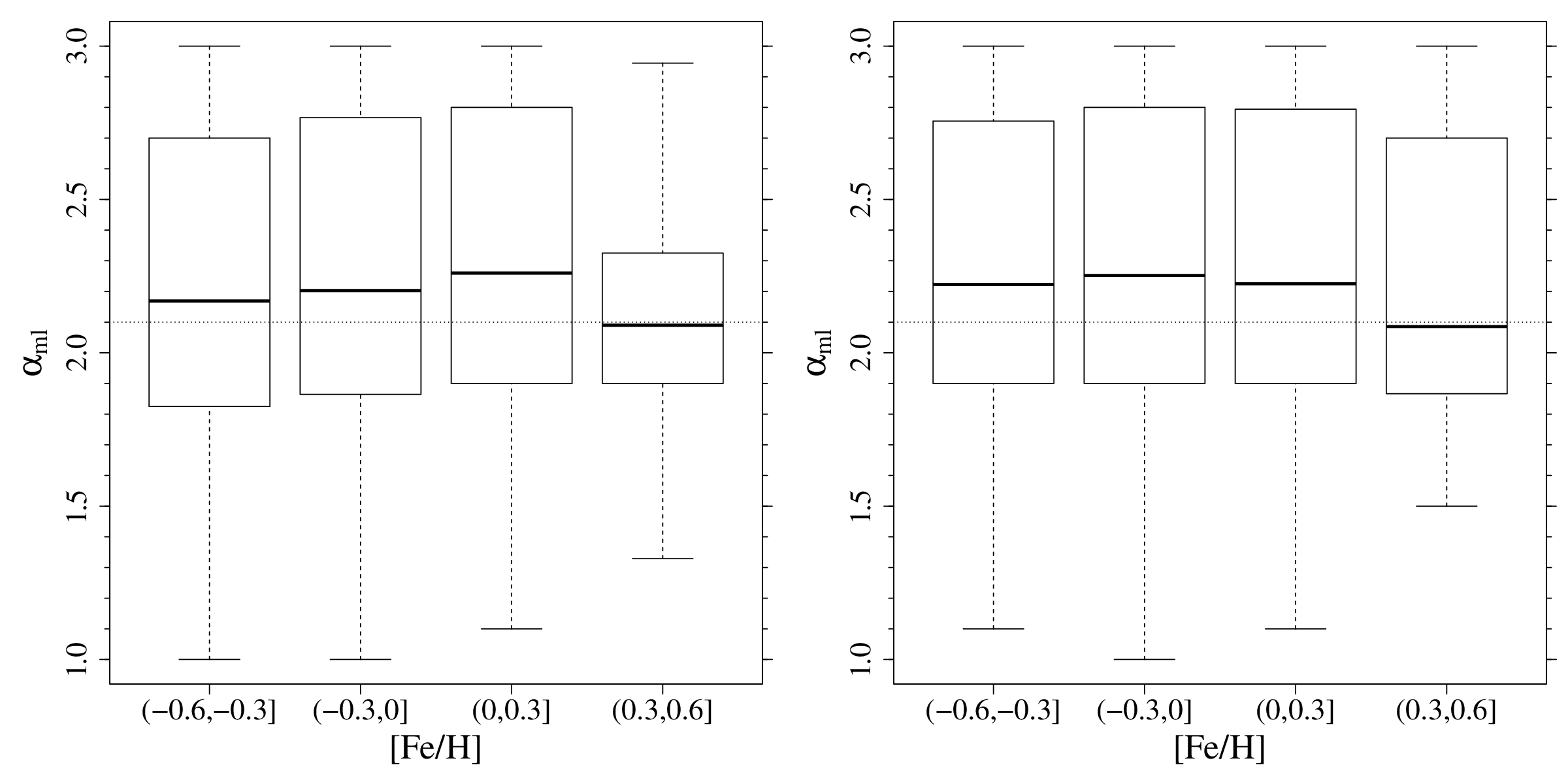}
        \caption{{\it Left}: Boxplot of the $\alpha_{\rm ml}$ sampled from models with modified boundary conditions, and recovered on the AS09 grid of models. Only MS models were considered. {\it Right}: Same as in the left panel, but adopted for RGB models.}
        \label{fig:effetto-BC}
\end{figure*}

\subsection{Effect of changing the boundary conditions}\label{sec:boundary}

Other possible sources of spurious trends are the input physics required to compute stellar models. As is well known, the location of an evolutionary track of given mass and chemical composition ($Z$ and $Y$) in the HR diagram or in the multidimensional hyperspace of the observables used in this paper depends on the adopted radiative opacity, equation of state, outer boundary conditions, etc. As a consequence, a change in the input physics necessarily affects the recovered stellar parameters \citep[see e.g.][]{Basu2012, scepter1}. 

As a final test (scenario AS09-bc), we explored the effect of changing the outer boundary conditions on the mixing-length estimate. We analysed a non-standard scenario in which the artificial stars and grid of models adopt different outer boundary conditions. In greater detail we built a new mock dataset sampled from a new non-standard grid of models computed with $\Delta Y/\Delta Z = 2.0$ but adopting the outer boundary conditions from non-grey atmosphere models. In particular we used the detailed atmospheric calculations by \citet{brott05}, computed using the PHOENIX code \citep{hauschildt99,hauschildt03},  available in the range $3000 \; {\rm K} \le T_{\rm eff} \le 10000 \; {\rm K}$, $0.0 \le \log g \; {\rm (cm \; s^{-2})} \le 5.0$, and $-4.0 \le {\rm[M/H]} \le 0.5$ where adopted. In the range $10000 \; {\rm K} \le T_{\rm eff} \le 50000 \; {\rm K}$, $0.0 \le \log g \; {\rm (cm \; s^{-2})} \le 5.0$, and $-2.5 \le {\rm [M/H]} \le 0.5$, where models from \citet{brott05} are unavailable, models by \citet{castelli03} are used. 

As is well known, a change in the outer boundary conditions  also affects the solar calibration of the mixing-length \citep[see e.g.][]{Montalban2004, Tognelli2011}. With the adopted non-grey boundary conditions, the new solar calibrated value of the mixing-length parameter is $\alpha_{\rm ml} = 1.74$. We used this value to build the non-standard mock dataset. We then applied to this mock catalogue the SCEPtER pipeline relying on the standard AS09 grid of stellar models, which has been computed adopting as outer boundary conditions by integrating the $T(\tau)$ relation by \citet{KrishnaSwamy1966}.

Fig.~\ref{fig:effetto-BC} shows that the reconstructed $\alpha_{\rm ml}$  are slightly overestimated with respect to the solar calibrated value of the recovery grid, and no particular differences exist between MS and RGB evolutionary stages. As a comparison, Sect.~\ref{sec:errori}  shows that for the AS09 scenario an increasing trend of $\alpha_{\rm ml}$ with [Fe/H] is expected in the RGB phase, while the recovery in the MS phase is nearly identical to that in the left panel of Fig.~\ref{fig:effetto-BC}. 

The existence of a difference only in the RGB phase is easily understood because the solar calibration of the mixing-length value makes the two set of tracks (standard and modified boundary conditions) nearly indistinguishable in the MS phase, while large differences exist in the RGB phase.

\section{Conclusions}\label{sec:conclusions}

We critically analysed the soundness of inferring the metallicity dependence of the mixing-length parameter from field star observations. In particular we discussed the theoretical foundations and the statistical reliability of the mixing-length calibration by means of standard ($T_{\rm eff}$ and [Fe/H]) and global asteroseismic ($\Delta \nu$ and $\nu_{\rm max}$) observables of field stars.

We followed a purely theoretical approach that allows us to perfectly control and disentangle the various error sources and  the systematic biases. We used a mock dataset of artificial stars rather than a catalogue of real objects. The advantage of this approach is that we exactly know all the characteristics of the artificial stars as we directly computed them. Then we applied to the mock catalogue the same maximum-likelihood procedure usually used to infer the parameters of observed stars, mixing-length included. The comparison between the recovered and true mixing-length values allowed us to quantify the random errors caused by the observational uncertainties, and the systematic biases due to different reasons.

We performed the analysis in different configurations. The starting point was a scenario in which the mock stars and stellar models used in the recovery are in perfect agreement with each other (scenario AS09). This represents the best possible scenario and allowed us to quantify the minimum random errors and biases affecting the estimated mixing-length $\alpha_{\rm ml}$. 

Then we explored a scenario in which the heavy-element mixture used to compute the stellar models for the recovery was different from that adopted in the artificial stars. We chose for the former the solar mixture by \citet{AGSS09}, i.e. AS09 scenario, and for the latter the \citet{GS98} mixture, i.e. GS98 scenario.
In all the considered scenarios the artificial stars had the same constant solar-calibrated $\alpha_{\rm ml} = 2.1$. 

As a first important result we found a huge spread in the estimated $\alpha_{\rm ml}$ values, which practically covered the entire allowed range (from 1 to 3). Both the AS09 and GS98 scenarios showed an overestimation  bias and the recovered mean values were $2.20 \pm 0.52$ and $2.24 \pm 0.48$, respectively. 
These results are of particular interest, having a relevant impact on any investigation that attempts a calibration of $\alpha_{\rm ml}$ from field stars. The very low precision attainable suggests that the procedure is poorly reliable for single star fits, at least as far as standard and global asteroseismic observables are available. Therefore calibration attempts of the mixing-length value based on field stars should be considered cautiously. Moreover, even after reducing the assumed observational errors by a factor of four, the variability on the recovered values was still impressively large, being one-half of that from standard scenario.
A much better result from empirical calibrations can be reached by relying on cluster stars \citep[see e.g.][]{Palmieri2002, Yildz2008, Basu2010, cluster2018}. In this case it is possible to reduce the error in the estimates largely by assuming a common chemical composition and a common age for all the stars.            
 
It is worth noting that the large spread of the estimated $\alpha_{\rm ml}$ values for the AS09 scenario is only due to the propagation of random observational errors; all the input physics of mock data and recovery grid of models perfectly match. It is however clear that this assumption is largely overly optimistic when considering real observational objects. In this case the presence of systematic discrepancies can lead either to a non-negligible offset in the recovered parameter and possibly to spurious trends.  

Indeed the GS98 scenario, which assumes a different reference solar heavy-element mixture between the mock dataset and recovery stellar grid, showed a strong presence of a trend in the fitted $\alpha_{\rm ml}$ versus the observed [Fe/H]. Such a trend is statistically robust, nevertheless it is completely spurious since the artificial stars have been computed at constant mixing-length $\alpha_{\rm ml} = 2.1$. In other words, a wrong assumption about the heavy-element mixture adopted in the models used to constrain the mixing-length parameter leads to an artefact.  

We proved that this spurious trend of the estimated $\alpha_{\rm ml}$ with metallicity is caused by the discrepancy between the $Y$ -- $Z$ -- [Fe/H] relation adopted to compute stellar models and that used for artificial stars. In fact, a change in the heavy-element mixture alters the $Y$ -- $Z$ -- [Fe/H] relation. As a consequence, the same [Fe/H] corresponds to different $Z$ or $Y$ values in models and stars. A mismatch in the $Y$ -- $Z$ -- [Fe/H] relation induces in turn a systematic discrepancy between models and stars, which is a function of the metallicity. Thus in any maximum-likelihood procedure, such a discrepancy must be counterbalanced by  changing the mixing-length parameter of the best-fitted model. The extent of the induced change is in turn a function of metallicity. As a consequence a spurious trend in the inferred $\alpha_{\rm ml}$ with [Fe/H] arises. Since the $Y$ -- $Z$ -- [Fe/H] relation can be altered not only by changing the heavy-element mixture but also modifying the helium-to-metal enrichment ratio $\Delta Y/\Delta Z$, another cause of spurious trend in the inferred $\alpha_{\rm ml}$ with [Fe/H] is the discrepancy between the $\Delta Y/\Delta Z$ value adopted in the recovery models and in mock stars.

We note that this artefact is the result of topological behaviour of stellar tracks in the hyperspace of parameters ($T_{\rm eff}$, [Fe/H], $\Delta \nu$, $\nu_{\rm max}$) and is therefore intrinsic and not related to the details of the fitting algorithm.

We directly verified that a trend similar to that in the GS98 scenario appears when the mock dataset and the grid of models share the same heavy-element mixture (AS09) but differs for the initial helium abundance. We studied the impact of assuming a wrong helium-to-metal enrichment ratio $\Delta Y/\Delta Z$ in the grid of models used in the recovery. Owing to the dependence of the [Fe/H] -- $Z$ relation on the assumed $\Delta Y/\Delta Z,$ it is possible to obtain largely different trends in the recovery. When $\Delta Y/\Delta Z = 1.0$ was assumed in the recovery (instead of the true value of 2.0) a large increasing trend of $\alpha_{\rm ml}$ with metallicity was obtained, even for AS09 scenario. On the other hands, for $\Delta Y/\Delta Z = 3.0$ the fitting process recovered a slightly decreasing trend for AS09 scenario. The same qualitative behaviours were detected for GS98 scenario.

Quite interestingly, we also found that a systematic error in the iron abundance scale -- mimicked by a rigid offset in the [Fe/H] of the artificial stars -- induces a spurious trend with [Fe/H] in the inferred $\alpha_{\rm ml}$ similar to that of GS98 scenario. We artificially reduced the [Fe/H] of the AS09 mock catalogue by $-0.1$ dex in [Fe/H], which nearly corresponds to the difference between GS98 and AS09 solar mixtures, and then we applied the standard fitting procedure. A clear trend of about one-half of the GS98 solar mixture was detected in this case. Therefore it seems that even a simple metallicity offset is a relevant cause for the trend development.

The  presence and origin of the observed trends with metallicity obviously raises some additional doubts concerning analyses of field stars with poorly determined chemical abundances.  Optimistically, we can  hope that no systematic is present on the [Fe/H] determinations; in this case the law of large numbers suggests that the effect of random errors in the metallicity determinations will on average compensate because overestimations and underestimations equally occur. 
However, the law of large numbers will not help in averaging out the biases on the not well-determined original helium abundance or chemical mixture. Moreover it will also not help to average out other systematic discrepancies between the observed stars and the computed evolutionary models. In particular, different assumptions about the boundary conditions,  equation of state, and the mean Rosseland opacity will lead to a change in the stellar track morphology. The actual relevance of some of these uncertainty sources on the reconstructed $\alpha_{\rm ml}$ and on the possible development of spurious trends is still unexplored. 

In this framework, we verified the relevance of a wrong assumption on the microscopic diffusion efficiency, effective temperature scale, and atmospheric boundary conditions. Neglecting the modification of the surface [Fe/H] caused by the microscopic diffusion obviously alters the portion of the grid hyperspace adopted in the recovery by modifying the [Fe/H] to $Z$ relation. This relation is neither constant nor  monotonic with time because the effect of the microscopic diffusion on the surface [Fe/H] reaches a maximum at about 80\% of the MS stellar lifetime and also varies with the stellar mass and initial metallicity. As expected in light of the previous discussion, the results from this scenario showed the development of an artificial trend in the estimated $\alpha_{\rm ml}$ with [Fe/H].  
On the other hand, the constant modification of the effective temperature scale was compensated by a rigid offset in the recovered $\alpha_{\rm ml}$, without the development of any trend.
The adoption of different boundary conditions had a negligible impact in the MS, while producing a different behaviour with respect to AS09 scenario in the RGB phase. This result stems from the fact that the change of the boundary conditions was coupled with a change in the solar calibrated $\alpha_{\rm ml}$, resulting in small differences during the MS evolution, but large discrepancies in later evolutionary stages.

The present research only scrapes the surface of the possible effects of the input physics mismatches between the mock data and recovery models on the $\alpha_{\rm ml}$ calibration. Further theoretical investigations with the aim to answer to these questions are highly desirables.

In summary, it seems that the estimate of the mixing-length parameter from single fields stars, for which only standard and global asteroseismic observables are available, is very problematic, even in the ideal case where the potential error sources in the input physics are taken under control. Moreover, a trend in the estimated $\alpha_{\rm ml}$ values versus the metallicity can originate from simple causes, such as an offset in the [Fe/H] scale or a wrong assumption of the $\Delta Y/\Delta Z$ or heavy-element mixture values. Because of the obvious difficulties in obtaining precise and accurate abundance measurements for large number of fields stars, their adoption for $\alpha_{\rm ml}$ calibration purposes is somewhat questionable. 

Ultimately, the adopted observational constraints proved to be not stringent enough to correctly disentangle the effect of the metallicity and mixing-length value on the stellar evolutionary tracks; as such, any claim about the possible dependence of the inferred mixing-length parameter on the metallicity for field stars should be considered cautiously.

\begin{acknowledgements}
We thank our anonymous referee for useful comments and suggestions that helped us clarify the manuscript.
This work has been supported by PRA Universit\`{a} di Pisa 2018-2019 
(\emph{Le stelle come laboratori cosmici di Fisica fondamentale}, PI: S. Degl'Innocenti) and by INFN (\emph{Iniziativa specifica TAsP}).
\end{acknowledgements}

\bibliographystyle{aa}
\bibliography{biblio}

\end{document}